\documentstyle[aps]{revtex} 
\input amssym.def 
\input amssym.tex 

\newcommand{\kasten}{\,\rule{1ex}{1ex}}

\def\m@th{\mathsurround=0pt}
\def\mymatrix#1{\null\,\vcenter{\normalbaselines\m@th
\ialign{\hfil$\scriptstyle ##$\hfil&&\quad\hfil $\scriptstyle
##$\hfil\crcr \mathstrut\crcr\noalign{\kern-\baselineskip}
#1\crcr\mathstrut\crcr\noalign{\kern-\baselineskip}}}\,}

\begin{document} 
\draft

\title{Riemannian symmetric superspaces and their origin in
random-matrix theory}

\author{Martin R. Zirnbauer${}^*$}

\address{Institute for Theoretical Physics, UCSB, Santa Barbara, U.S.A.}

\date{March 31, 1996}
\maketitle

\begin{abstract}
Gaussian random-matrix ensembles defined over the tangent spaces of
the large families of Cartan's symmetric spaces are considered.  Such
ensembles play a central role in mesoscopic physics as they describe
the universal ergodic limit of disordered and chaotic single-particle
systems.  The generating function for the spectral correlations of
each ensemble is reduced to an integral over a Riemannian symmetric
superspace in the limit of large matrix dimension.  Such a space is
defined as a pair $(G/H,M_{\rm r})$ where $G/H$ is a complex-analytic
graded manifold homogeneous with respect to the action of a complex
Lie supergroup $G$, and $M_{\rm r}$ is a maximal Riemannian
submanifold of the support of $G/H$.
\end{abstract}

\bigskip
${}^*$ Permanent address: Institut f\"{u}r Theoretische Physik, 
Universit\"{a}t zu K\"{o}ln, Germany

\section{Introduction}

The mathematics of supersymmetry, though conceived and developed in
elementary particle theory, has been applied extensively to the
physics of disordered metals during the past decade.  Improving on
earlier work by Wegner\cite{wegner,sw}, Efetov\cite{efetov} showed how
to approximately map the problem of calculating disorder averages of
products of the energy Green's functions for a single electron in a random
potential, on a supersymmetric nonlinear $\sigma$ model.  Later it was
shown\cite{vwz} that the same nonlinear $\sigma$ model describes the
large-$N$ limit of a random-matrix ensemble of the
Wigner-Dyson\cite{dyson} type.  Since then, Efetov's method has
evolved into a prime analytical tool in the theory of disordered or
chaotic mesoscopic single-particle systems.  Competing methods are
limited either to the diffusive regime (the impurity diagram
technique), or to isolated systems in the ergodic regime (the
Dyson-Mehta orthogonal polynomial method), or to quasi-one-dimensional
systems (the DMPK equation).  In contrast, Efetov's method is
applicable to isolated and to open systems in the diffusive, ergodic,
localized, and even ballistic regime, to both spectral correlations
and transport properties, and it can in principle be used in any
dimension.  This versatility has engendered a large body of nontrivial
applications, many of which are outside the range of other methods.
Of these, let me mention: (i) the Anderson transition on a Bethe
lattice\cite{efetov84,mrz85,mf91}, (ii) localization in disordered
wires\cite{el,mrz91,mrz92,mmz,fb}, (iii) multifractality of energy
eigenstates in two dimensions\cite{mk,ef,mirlin95}, (iv) weak
localization and conductance fluctuations of chaotic billiards
strongly coupled to a small number of scattering
channels\cite{heidelberg,zukforr} and, most recently, (v) a
theoretical physicist's proof of the Bohigas-Giannoni-Schmit
conjecture for chaotic Hamiltonian systems\cite{mk1,aasa}.

In spite of these manifest successes, Efetov's supersymmetry method
has been ignored (for all that I know) by mathematical physicists.
This is rather unfortunate for several reasons.  First, an infusion of
mathematical expertise is needed to sort out some matters of principle
and promote the method to a rigorous tool.  Second, various extensions
of currently available results seem possible but have been hindered by
the lack of mathematical training on the part of the condensed matter
theorists applying the method.  And third, the geometric structures
underlying Efetov's nonlinear $\sigma$ models are of exquisite beauty
and deserve to be studied in their own right.  Part of the reason why
neither mathematicians nor mathematical physicists have monitored or
contributed to the development, may be that there does not exist a
concise status report that would appeal to a mind striving for clarity
and rigor.  Hence the first, and very ambitious, motivation for 
getting started on the present paper was to make an attempt and partially
fill the gap.

Another objective is to report on a recent extension of the
supersymmetry method to random-matrix theories beyond the standard
Wigner-Dyson ones.  In her study of Anderson localization in the
presence of an A-B sublattice symmetry, Gade\cite{gade} noticed that
the manifold of the nonlinear $\sigma$ model is promoted to a larger
manifold at zero energy.  The same phenomenon occurs in the chiral
limit of the QCD Dirac operator at zero virtuality\cite{jjmv}.  For
several years it remained unclear how to handle this enlargement of
the manifold in the supersymmetric scheme.  (Gade used the replica
trick instead of supersymmetry.)  The key to solving the problem can
be found in a paper by Andreev, Simons, and Taniguchi\cite{ast} who
observed that what one needs to do is to avoid complex conjugation
of the anticommuting variables.  In the present paper I will elaborate
on this observation and cast it in a concise mathematical language.
Moreover, I will show that the same technical innovation allows to
treat the random-matrix theories that arose\cite{az_prl,az96} in the
stochastic modeling of mesoscopic metallic systems in contact with a
superconductor.

An outline of the basic mathematical structure is as follows.
Consider a homogeneous space $G/H$, where $G$ and $H$ are complex Lie
supergroups, and regard $G/H$ as a complex-analytic
$(p,q)$-dimensional supermanifold in the sense of
Berezin-Kostant-Leites\cite{bl,kostant}.  To integrate its holomorphic
sections, select a closed, oriented and real $p$-manifold $M_{\rm r}$
contained in the support $M = G_0 / H_0$ of the supermanifold.  The
natural (invariant) supergeometry of $G/H$ induces a geometry on
$M_{\rm r}$ by restriction.  If this geometry is Riemann and $M_{\rm
r}$ is a symmetric space, the pair $(G/H,M_{\rm r})$ is called a
Riemannian symmetric superspace.  This definition will be shown to be
the one needed for the extension of the supersymmetry method beyond
Wigner-Dyson.  The difficulties disordered single-particle theorists
had been battling with were caused by the fact that the exact sequence
        \[
        0 \rightarrow {\rm nilpotents} \rightarrow G/H \rightarrow
        M \rightarrow 0
        \]
does not, in general, reduce to an exact sequence of sheaves of
{\it real-analytic} sections terminating at the Riemannian submanifold
$M_{\rm r}$.

When integrating the invariant holomorphic Berezin superform on $G/H$,
one must pay careful attention to its coordinate ambiguity.  This
subtle point is reviewed in Sec.~\ref{sec:superintegration}.  After a
brief reminder of the procedure of Grassmann-analytic continuation (in
Sec.~\ref{sec:g_a_c}), the complex Lie supergroups ${\rm Gl}(m|n)$ and
${\rm Osp}(m|2n)$ (in Sec.~\ref{sec:superLie}), and Cartan's symmetric
spaces (in Sec.~\ref{sec:sym_spaces}), the details of the definition
of Riemannian symmetric superspaces are given in Sec.~\ref{sec:r_s_s}.
Table 2 lists the large families of these spaces.

Sec.~\ref{sec:III}, the largest of the paper, treats the Gaussian
random-matrix ensemble defined over the symplectic Lie algebra ${\rm
sp}(N)$, by an adaptation of Efetov's method.  A simple example
(Sec.~\ref{sec:example}) illustrates the general strategy.  Details of
the method, including a complete justification of all manipulations
involved, are presented in Secs.~\ref{sec:def_C} --
\ref{sec:saddle_points}.  Theorem 3.3 expresses the Gaussian ensemble
average of a product of $n$ ratios of spectral determinants as a
superintegral.  Theorem 3.4 reduces this expression to an integral
over the Riemannian symmetric superspace ${\rm Osp}(2n|2n) / {\rm
Gl}(n|n)$ with $M_{\rm r} = \left( {\rm SO}^*(2n) / {\rm U}(n) \right)
\times \left( {\rm Sp}(n) / {\rm U}(n) \right)$, in the limit $N \to
\infty$.

According to Cartan's list, there exists eleven large families of
symmetric spaces.  Ten of these correspond to universality classes
that are known to describe disordered single-particle systems in the
ergodic regime\cite{az_prl,az96}.  The class singled out for detailed
treatment in Sec.~\ref{sec:III} describes mesoscopic
normal-superconducting hybrid systems with time-reversal symmetry
broken by a weak magnetic field.  The remaining nine classes are
briefly discussed in Sec.~\ref{sec:IV}.  Each of them is related, by
the supersymmetry method, to one of the large families of Riemannian
symmetric superspaces of Table 2.  A summary is given in
Sec.~\ref{sec:conclusion}.

\section{Riemannian symmetric superspaces}

\subsection{The Berezin integral on analytic supermanifolds}
\label{sec:superintegration}

Let $A(U)$ denote the algebra of analytic functions on an open subset
$U$ of $p$-dimensional real space.  By taking the tensor product with
the Grassmann algebra with $q$ generators one obtains $A(U) \otimes
\Lambda({\Bbb R}^q)$, the algebra of analytic functions on $U$ with
values in $\Lambda({\Bbb R}^q)$.  Multiplication on $\Lambda( {\Bbb
R}^q)$ is the exterior one, so the algebra is supercommutative (or
graded commutative).  The object at hand serves as a model for what is
called a real-analytic $(p,q)$-dimensional supermanifold (or graded
manifold\cite{bbh}) in the sense of Berezin, Kostant, and Leites
(BKL)\cite{bl,kostant}; which, precisely speaking, is a sheaf of
supercommutative algebras ${\cal A}$ with an ideal ${\cal N}$ (the
nilpotents), such that ${M} \simeq {\cal A}/{\cal N}$ is an analytic
$p$-manifold and on a domain $U \subset M$, ${\cal A}$ splits as $A(U)
\otimes \Lambda({\Bbb R}^q)$.  The global sections of the bundle
${\cal A} \to {M}$ are called superfunctions, or functions for short.
${M}$ is called the underlying space, or base, or support, of the
supermanifold.  $M$ will be assumed to be orientable and closed
$(\partial M = 0)$.

The calculus on analytic supermanifolds is a natural extension of the
calculus on analytic manifolds.  Functions are locally expressed in
terms of (super-)coordinates $(x;\xi) := (x^1, ..., x^p; \xi^1, ...,
\xi^q)$ where $x^i$ $(\xi^j)$ are even (resp. odd) local sections of
${\cal A}$.  If $(x;\xi)$ and $(y;\eta)$ are two sets of local
coordinates on domains that overlap, the transition functions $y^i =
f^i(x;\xi)$ and $\eta^j = \varphi^j(x;\xi)$ are analytic functions of
their arguments and are consistent with the ${\Bbb Z}_2$-grading of
${\cal A}$.

In what follows the focus is on the theory of integration on analytic
supermanifolds.  Recall that on $p$-manifolds the objects
one integrates are $p$-forms and their transformation law is given by
        \[ 
        {\rm d}y^1 \wedge ... \wedge {\rm d}y^p = {\rm d}x^1 \wedge
        ... \wedge {\rm d}x^p \ {\mathop{\rm Det}\nolimits} \left( 
        {\partial y^i /  \partial x^j} \right) .  
        \] 
The obvious (super-)generalization of the Jacobian 
${\mathop{\rm Det}\nolimits} \left( {\partial y^i /  \partial x^j} 
\right)$ is the Berezinian\cite{berezin}
        \[ 
        {\rm Ber} \left( {y,\eta / x,\xi}
        \right) := {\mathop{\rm SDet}\nolimits} \pmatrix{ { \partial y^i /
        \partial x^j } &{ \partial y^i / \partial \xi^j }\cr 
        { \partial \eta_i / \partial x^j} &{ \partial \eta_i /
        \partial \xi^j }\cr} , 
        \] 
where ${\mathop{\rm SDet}\nolimits}$ is the symbol for superdeterminant.  
Guided by analogy, one postulates that an integral superform ought to be 
an object $\tilde D$ transforming according to the law
        \begin{equation}
        \tilde D(y,\eta) = \tilde D(x,\xi) \ {\rm Ber} 
        \left( { y,\eta / x,\xi } \right) .  
        \label{berezin_law} 
        \end{equation} 
A natural candidate would seem to be 
        \[ 
        D(x,\xi) := {\rm d}x^1 \wedge ... \wedge {\rm d}x^p 
        \otimes \partial_{\xi^1} ... \partial_{\xi^q} ,
        \] 
which is a linear differential operator taking superfunctions $f$ into
$p$-forms $D[f]$ ($\partial_{\xi^i}$ denotes the partial derivative
with respect to the anticommuting coordinate $\xi^i$).  The $p$-form
$D[f]$ can be integrated in the usual sense to produce a number.
However, the transformation law for $D(x,\xi)$ turns out to be
not quite (\ref{berezin_law}) but rather
        \begin{equation}
        D(y,\eta) = D(x,\xi) \ {\rm Ber} \left( 
        { y,\eta / x,\xi } \right) + \beta .  
        \label{anomalous_law}
        \end{equation}
An explicit description of the term $\beta$ on the right-hand side,
here referred to as the {\it anomaly}, was first given by
Rothstein\cite{rothstein2}.  It is nonzero whenever some even
coordinate functions are shifted by nilpotent terms.  Its main
characteristic is that on applying it to a superfunction $f$, one gets
a $p$-form that is {\it exact}: $\beta[f] = {\rm d} (\alpha[f])$.

The existence of an anomaly in the transformation law for $D(x,\xi)$
leads one to consider a larger class of objects, namely $\Lambda^p (M)
\otimes {}_{\cal A} {\cal D}$, the sheaf of linear differential
operators on ${\cal A}$ with values in the $p$-forms on ${M}$.
($\Lambda^p({M}) \otimes {}_{\cal A}{\cal D}$ naturally is a right
${\cal A}$-module.)  To rescue the simple transformation law
(\ref{berezin_law}) one usually passes from $\Lambda^p({M}) \otimes
{}_{\cal A} {\cal D}$ to its quotient by the
anomalies\cite{rothstein2}.  In order for the integral to be
well-defined over the quotient, one must take the functions one
integrates to be compactly supported.

Sadly, this last option {\it is not available to us}.  The functions
that will be encountered in the applications worked out below,
do not ever have compact support but are {\it analytic} functions
instead.  When integrating such functions, we need to work with the
full transformation law (\ref{anomalous_law}), which includes the
anomaly.

Another way of avoiding the anomaly is to arrange for the transition
functions never to shift the even coordinates by nilpotents, by
constructing a restricted subatlas\cite{rogers3}.  However, because
the concept of a restricted subatlas is somewhat contrived, this
approach has been found to be of limited use in the type of problem
that is of interest here.

To arrive at a definition of superintegration that is useful in
practice, we proceed as follows.  The supermanifold is covered by a
set of charts with domains $U_i$ and coordinates $(x_{(i)},\xi_{(i)})$
($i=1,...,n$).  On chart $i$ let $\omega_i := D(x_{(i)}, \xi_{(i)})
\circ \tilde\omega_i$ with $\tilde\omega_i$ a local section of ${\cal
A}$, and let $\alpha_i \in \Lambda^{p-1}(M) \otimes {}_{\cal A}{\cal
D}|_{U_i}$.  Partition ${M}$ into a number of consistently oriented
$p$-cells $D_1, ..., D_n$, with $D_i$ contained in $U_i$.  For $i < j$
put $D_{ij} := \partial D_i \cap \partial D_j$ and, if $D_{ij}$ is
nonempty and is a $(p-1)$-cell, fix its orientation by $\partial D_i =
+ D_{ij} +...~$.

{\it Definition 2.1}:  A collection $\{ \omega_i , \alpha_i \}_{i=1, 
...,n}$ is called a {\it Berezin measure} $\omega$ if the conditions
        \begin{eqnarray} 
        &&\tilde\omega_i / \tilde\omega_j = {\rm Ber}( i / j ) ,
        \label{consistent_1} \\ 
        &&\omega_i + {\rm d}\alpha_i = \omega_j + {\rm d}\alpha_j ,
        \label{consistent_2}
        \end{eqnarray} 
are satisfied on overlapping domains.  The Berezin integral $f \mapsto
\int_{M} \omega[f]$ is defined
        \begin{equation}
        \int_{M} \omega[f] = \sum_{i=1}^n \int_{D_i} \omega_i [f] +
        \sum_{i < j} \int_{D_{ij}} \alpha_{ij}[f]
        \label{berezin_integral}
        \end{equation}
where $\alpha_{ij} = \alpha_i - \alpha_j$.  The quantities $\omega_i$
and $\alpha_i$ are called the principal term and the anomaly of the
Berezin measure on chart $i$.

{\it Remark 2.2}: The conditions (\ref{consistent_1}) and
(\ref{consistent_2}) ensure the existence of a global section $\omega
\in \Lambda^p({M}) \otimes {}_{\cal A}{\cal D}$ whose local expression
in chart $i$ is $\omega_i + {\rm d}\alpha_i$.  The existence of
$\omega$ means that the distribution (\ref{berezin_integral}) is
independent of the coordinate systems and the cell partition chosen.
Because (\ref{berezin_integral}) depends only on the differences
$\alpha_i - \alpha_j$, one can gauge the anomaly to zero on one of the
charts without changing the Berezin integral.

{\it Example 2.3}: Consider the real supersphere ${\rm S}^{p|2}$, a
$(p,2)$-dimensional supermanifold with support ${\rm S}^p$, which is
the space of solutions in $(p+1,2)$ dimensions of the quadratic equation
        \[
        {\tilde x}_0^2 + {\tilde x}_1^2 + ... + {\tilde x}_p^2 
        + 2{\tilde \xi}_1 {\tilde \xi}_2 = 1 .
        \]
Cover ${\rm S}^p$ by two domains 1 and 2 obtained by removing
the south $(\tilde x_0 = - 1)$ or north pole $(\tilde x_0 = + 1)$.  
Introduce stereographic coordinates
$(x_1,...,x_p;\xi_1,\xi_2)$ and $(y_1,...,y_p;\eta_1,\eta_2)$ for
${\rm S}^{p|2}$ on these domains with transition functions
        \[
        y_1 = - x_1 / R^2 , \quad
        y_i = x_i / R^2 \ (i = 2, ..., p), \quad
        \eta_j = \xi_j / R^2 \ (j = 1, 2)
        \]
where $R^2 = \sum_{i=1}^p x_i^2 + 2\xi_1\xi_2$.  (The minus sign
preserves the orientation.)  Consider 
        \begin{eqnarray}
        \omega_1 &=& D(x,\xi) \circ \left( 1 + {\textstyle\sum} x_i^2
        + 2 \xi_1 \xi_2 \right)^{-p+2} ,
        \nonumber \\
        \omega_2 &=& D(y,\eta) \circ \left( 1 + {\textstyle\sum} y_i^2
        + 2 \eta_1 \eta_2 \right)^{-p+2} ,
        \nonumber \\
        \alpha_{12} &=& - \Omega { (\sum x_i^2 )^{(p-2)/2} \over
        ( 1 + \sum x_i^2 )^{p-2}} \otimes 2 \partial_{\xi_1}
        \partial_{\xi_2} \circ \xi_1 \xi_2 ,
        \nonumber
        \end{eqnarray}
where $\Omega = (\sum_j x_j^2 ) ^{-p/2} \sum_{i=1}^p (-1)^i {\rm d}x_1
\wedge ... \wedge {\rm d}x_{i-1} \wedge x_i {\rm d}x_{i+1} \wedge
... \wedge {\rm d}x_p$ is the solid-angle $(p-1)$-form in $p$
dimensions.  It is not difficult to check by direct calculation that
$\omega_1$, $\omega_2$ and $\alpha_1 = \alpha_{12}$, $\alpha_2 \equiv
0$ obey the relations (\ref{consistent_1}) and (\ref{consistent_2}).
Hence, they express a globally defined Berezin measure $\omega$ in the
sense of Definition 2.1.  (The geometric meaning of $\omega$ will be
specified in Sec.~\ref{sec:superLie}.)  For $p \ge 3$, the anomaly
$\alpha_{12}$ scales to zero when $\sum x_i^2 \to \infty$, so we
may shrink cell 2 to a single point (a set of measure zero) and
compute the Berezin integral simply from
        \[
        \int_{{\rm S}^p} \omega[f] 
        = \int_{{\Bbb R}^p} D(x,\xi) \left( 1 + {\textstyle\sum} x_i^2 
        + 2\xi_1 \xi_2 \right)^{-p+2} f(x;\xi) .
        \]
In these cases we can get away with using only a single chart.  The
situation is different for $p = 2$ and $p = 1$.  In the first case the
anomaly is scale-invariant (the solid angle is) and by again shrinking
cell 2 to one point (the south pole $(y_1,y_2) = (0,0)$ on ${\rm
S}^2$) we get
        \[
        \int_{{\rm S}^2} \omega[f] = \int_{{\Bbb R}^2} D(x,\xi)
        f(x;\xi) + 4\pi f \Big|_{\rm south~pole} .
        \]
In particular, $\int_{{\rm S}^2} \omega[1] = 4\pi$.  For $p = 1$ the
anomaly diverges at $x = 0$ and $x = \infty$.  In this case the
general formula (\ref{berezin_integral}) must be used, and one finds
$\int_{{\rm S}^1} \omega[1] = 0$.

\subsection{Grassmann-analytic continuation}
\label{sec:g_a_c}

In the formulation of BKL, the vector fields of a supermanifold do
not constitute a module over ${\cal A}$ but are constrained to be {\it
even} derivations of ${\cal A}$, which is to say that their coordinate
expression is of the form
        \[
        \hat X = f^i(x;\xi) {\partial\over\partial x^i} + 
        \varphi^j(x;\xi) {\partial\over\partial\xi^j} 
        \]
where $f^i$ and $\varphi^j$ are even and odd superfunctions
respectively.  Unfortunately, this formulation is too narrow for most
purposes.  The reason is that in applications one typically deals not
with a single supermanifold but with many copies thereof (one per
lattice site of a lattice-regularized field theory, for example).  So
in addition to the anticommuting coordinates of the one supermanifold
that is singled out for special consideration, there exist many more
anticommuting variables associated with the other copies of
the supermanifold.  When the focus is on one supermanifold, these can
be considered as ``parameters''.  Often one wants to make
parameter-dependent coordinate transformations, leading to
coefficients $f_{i_1...i_n}(x)$ in the expansion $f(x;\xi) = \sum
f_{i_1...i_n}(x) \xi^{i_1} ... \xi^{i_n}$ that depend on extraneous
Grassmann parameters.  (For example, when the supermanifold is a Lie
supergroup, it is natural to consider making left and right translations $g
\mapsto g_L g g_R$.)  The upshot is that one wants to take ${\cal A}$
as a sheaf of graded commutative algebras not over ${\Bbb R}$ but over
some (large) parameter Grassmann algebra $\Lambda$ (the Grassmann
algebra generated by the anticommuting coordinates of the ``other''
supermanifolds).  Making this extension, which is called
``Grassmann-analytic continuation'' in\cite{berezin}, one is led to
consider the more general class of vector fields of the form
        \begin{equation}
        \hat X = f^i(x,\xi;\beta) {\partial\over\partial x^i} + 
        \varphi^j(x,\xi;\beta) {\partial\over\partial\xi^j} 
        \label{vector_fields}
        \end{equation}
where the symbol $\beta$ stands for the extra Grassmann parameters and
the dependences on these are such that $f^i$ and $\varphi^j$ continue
to be even and odd respectively (the ${\Bbb Z}_2$-grading of ${\cal
A}$ after Grassmann-analytic continuation is the natural one).

The vector fields (\ref{vector_fields}) still are even derivations of
the extended algebra.  One can go further by demanding that ${\rm
Der}{\cal A}$ be free over ${\cal A}$ and including the odd ones, too.
When that development is followed to its logical conclusion, one
arrives at Rothstein's axiomatic definition\cite{rothstein1} of
supermanifolds, superseding an earlier attempt by
Rogers\cite{rogers1,rogers2}.  Although there is no denying the
elegance and consistency of Rothstein's formulation, we are not going
to embrace it here, the main reason being that odd derivations will
not really be needed.  For the purposes of the present paper we will
get away with considering vector fields of the constrained form
(\ref{vector_fields}).

\subsection{The complex Lie supergroups ${\rm Gl}(m|n)$ and 
${\rm Osp}(m|2n)$} \label{sec:superLie}

The supermanifolds we will encounter all derive from the complex Lie
supergroups\cite{berezin,bbh} ${\rm Gl}(m|n)$ and ${\rm Osp}(m|2n)$,
by forming cosets.  The definition of ${\rm Gl}(m|n)$ rests on the
notion of an invertible supermatrix $g = \left( \mymatrix{g_{00}
&g_{01}\cr g_{10} &g_{11}\cr} \right)$ where $g_{00}$, $g_{01}$,
$g_{10}$ and $g_{11}$ are matrices of size $m\times m$, $m\times n$,
$n\times m$, and $n\times n$.  The supermanifold structure of ${\rm
Gl}(m|n)$ comes from taking the matrix elements of $g_{00}$ and
$g_{11}$ ($g_{01}$ and $g_{10}$) for the even (resp. odd) coordinates
on suitable domains of the base ${M} = {\rm Gl}(m,{\Bbb C}) \times
{\rm Gl}(n,{\Bbb C})$.  The Lie supergroup structure derives from the
usual law of matrix multiplication.

For $m \not= n$, it is common practice to split off from ${\rm
Gl}(m|n)$ the ${\rm Gl}(1)$-ideal generated by the unit matrix, so as
to have an irreducible Lie superalgebra\cite{kac,scheunert}.  For $m =
n$, which turns out to be the case of most interest here, one ends up
having to remove two ${\rm Gl}(1)$'s, one generated by the unit matrix
and the other one by the superparity matrix $\sigma = {\rm diag}( 1_n ,
-1_n)$.  And even then the Lie superalgebra is not irreducible in a
sense, for the Killing form ${\mathop{\rm STr}}~{\mathop{\rm
ad} \nolimits}(X) {\mathop{\rm ad}\nolimits}(Y)$ vanishes identically.
We therefore prefer to take ${\rm Gl}(m|n)$ as it stands (with no
ideals removed) and replace the Killing form by the invariant
quadratic form ${\rm B}(X,Y) = {\mathop{\rm STr}\nolimits} XY$, which
is nondegenerate in all cases (including $m = n$).
 
The complex orthosymplectic Lie supergroup ${\rm Osp}(m|2n)$ is
defined as a connected subgroup of ${\rm Gl}(m|2n)$ fixed
by an involutory automorphism $g \mapsto \hat\tau(g) = \tau {g^{-1}} ^
{\rm T} \tau^{-1}$, where $\tau$ is supersymmetric $(\tau = \tau^{\rm
T}\sigma = \sigma \tau^{\rm T})$.\footnote{Supertransposition ${\rm T}$
is an operation with the properties $(AB)^{\rm T} = B^{\rm T} A^{\rm T}$
and $A^{\rm TT} = \sigma A \sigma$.}  The support of ${\rm Osp}(m|2n)$ is
${\rm SO}(m,{\Bbb C}) \times {\rm Sp}(n,{\Bbb C})$.

The action of a Lie supergroup on itself by left and right
translations gives rise to right- and left-invariant vector fields.  A
Berezin measure on a Lie supergroup is said to be invariant, and is
called a Berezin-Haar measure, if its Lie derivatives\cite{berezin}
with respect to the invariant vector fields vanish.

Given a Lie supergroup $G$ and a subgroup $H$, the coset superspace
$G/H$ is defined by decreeing that the structure sheaf of the coset
superspace is a quotient of sheaves.  The action of $G$ on $G/H$ by
left translation gives rise to so-called Killing vector fields.  A
Berezin measure on $G/H$ is called invariant if its Lie derivatives
with respect to the Killing vector fields are zero.

If ${\rm Osp}_{\Bbb R}(m|2n)$ denotes the orthosymplectic supergroup
over the reals, the coset space ${\rm Osp}_{\Bbb R}(m+1|2n) / {\rm
Osp}_{\Bbb R}(m|2n)$ can be identified with the real supersphere ${\rm
S}^{m|2n}$.  The Berezin measure discussed in Example 2.3 is invariant
with respect to the action of ${\rm Osp}_{\Bbb R}(p+1|2)$ on ${\rm
S}^{p|2}$ and can be viewed as the ``volume superform'' of ${\rm
S}^{p|2}$.  Hence we can restate the results of that example as
follows: ${\rm vol}({\rm S}^{2|2}) := \int_{{\rm S}^2} \omega[1] =
4\pi$ and ${\rm vol}({\rm S}^{1|2}) := \int_{{\rm S}^1} \omega[1] =
0$.

\subsection{Holomorphic Berezin measures on complex-analytic 
supermanifolds}

To go from real-analytic supermanifolds to complex-analytic ones, one
replaces the structure sheaf ${\cal A}$ by a sheaf of graded
commutative algebras ${\cal H}$ over ${\Bbb C}$ such that ${M} \simeq
{\cal H} / {\cal N}$ is a complex manifold and ${\cal H}$ is locally
modeled by $H(U) \otimes \Lambda({\Bbb C}^q)$, where $H(U)$ is the
algebra of holomorphic functions on $U \subset M$.  The natural objects
to consider then are holomorphic superfunctions, i.e. global sections
of the bundle ${\cal H} \to {M}$.  In local coordinates $z^1, ...,
z^p; \zeta^1, ... , \zeta^q$ such sections are written $f(z;\zeta)$.
Grassmann-analytic continuation is done as before when needed.  A
Berezin measure on a complex-analytic $(p,q)$-dimensional
supermanifold is a linear differential operator $\omega$ that takes
holomorphic superfunctions $f$ into holomorphic $p$-forms $\omega[f]$
on ${M}$.  The statements made in Sec.~\ref{sec:superintegration}
about the anomalous transformation behavior of Berezin measures apply
here, too (mutatis mutandi).

To define Berezin's integral in the present context, one more piece of
data must be supplied, namely a {\it real} $p$-dimensional
submanifold ${M}_{\rm r} \subset {M}$ over which the holomorphic
$p$-form $\omega[f]$ can be integrated to produce a complex
number.  Thus, given $\omega$ and ${M}_{\rm r}$, Berezin's integral is
the distribution
        \begin{equation}
        f \mapsto \int_{{M}_{\rm r}} \omega[f] .
        \label{holom_int}
        \end{equation}

Let me digress and mention that this definition, natural and simple as
it is, was not ``discovered'' by the random-matrix and mesoscopic
physics community (including myself) until quite recently.  With one
notable exception\cite{ast}, all past superanalytic work on disordered
single-particle systems employed some operation of ``complex
conjugation'' of the Grassmann generators -- namely an adjoint of the
first or second kind\cite{berezin} -- to make the treatment of the
ordinary (``bosonic'') and anticommuting (``fermionic'') degrees of
freedom look as much alike as possible.  Presumably this was done
because it was felt that such egalitarian treatment is what is
required by the principle of ``supersymmetry''.  Specifically, a
reality constraint was imposed not just on the underlying space $M$
(fixing ${M}_{\rm r}$) but on the entire structure sheaf to reduce
${\cal H}$ to a sheaf of algebras over ${\Bbb R}$.  Although this
reduction can be done with impunity in some cases (namely the classic
Wigner-Dyson symmetry classes), it has turned out to lead to
insurmountable difficulties in others (the chiral and
normal-superconducting symmetry classes).  A major incentive of the
present paper is to demonstrate that the construction
(\ref{holom_int}) is in fact the ``good'' one to use for the
application of supermanifold theory to disordered single-particle
systems in general.  Although that construction may hurt the
physicists' aesthetic sense by ``torturing supersymmetry'', it should
be clear that we are not breaking any rules.  Recall that according to
Berezin, superintegration is a two-step process: first, the
Fermi-integral (i.e. differentiation with respect to the anticommuting
coordinates) is carried out, and it is only {\it afterwards} that the
ordinary (Bose-) integrals are done.  When the sequential nature of
the Berezin integral is taken seriously, there is no compelling reason
why one should ever want to ``complex conjugate'' a Grassmann
variable.  In the present paper, we take the radical step of
abandoning complex conjugation of Grassmann variables altogether.

{\it Example 2.4}: The simplest nontrivial example\cite{ast} is given
by ${\rm Gl}(1|1)$, the Lie supergroup of regular complex $2\times
2$ supermatrices $g = \left( \mymatrix{ a &\beta\cr \gamma &d\cr}
\right)$ with support ${M} = {\rm Gl}(1,{\Bbb C}) \times {\rm
Gl}(1,{\Bbb C})$.  The Berezin-Haar measure on ${\rm Gl}(1|1)$ is
$\omega = (2\pi i)^{-1} D(ad;\beta\gamma)$ where $D(ad;\beta\gamma) =
{\rm d}a \wedge {\rm d}d \otimes \partial_\beta \partial_\gamma$.
Solving the regularity conditions $a \not= 0$ and $d \not= 0$ by
parameterizing ${\rm Gl}(1|1)$ through its Lie algebra, $g = \exp
\left( \mymatrix{z_1 &\zeta_1\cr \zeta_2 &z_2 \cr} \right)$, one finds
        \begin{eqnarray}
        2\pi i \omega &=& D(z_1 z_2;\zeta_1 \zeta_2) \circ 
        {(z_1 - z_2)^2 \over (1-e^{z_1 - z_2}) (e^{z_2 - z_1}-1)} 
        \nonumber \\
        &-& \left( {\rm d} \ln (e^{z_1} - e^{z_2}) - 
        {(z_1 - z_2) ({\rm d}z_1 -{\rm d}z_2)
        \over (1-e^{z_1 - z_2})(e^{z_2 - z_1}-1)} \right) \otimes
        \partial_{\zeta_1} \partial_{\zeta_2} \circ \zeta_1 \zeta_2 .
        \label{gl11}
        \end{eqnarray}
Note that this expression is holomorphic in a neighborhood of the
origin $z_1 = z_2 = 0$.  The first term on the right-hand side is the
principal term, and the second one is the anomaly of $\omega$ in these
coordinates.  To integrate $\omega$, one might be tempted to choose
for ${M}_{\rm r}$ the ${\rm U}(1) \times {\rm U}(1)$ subgroup defined
by ${\rm Re}(z_1) = 0 = {\rm Re}(z_2)$.  However, since the rank-two
tensor ${\mathop{\rm STr}\nolimits}~{\rm d}g {\rm d}g^{-1} = {\rm d}a
{\rm d}a^{-1} - {\rm d}d {\rm d}d^{-1} + {\rm nilpotents} = - {\rm
d}z_1^2 + {\rm d}z_2^2 + ...$ is not Riemann on ${\rm U}(1) \times
{\rm U}(1)$, this will not be the best choice.  A Riemannian structure
is obtained by taking ${M}_{\rm r} = {\Bbb R}^+ \times {\rm S}^1$
defined by ${\rm Im}(z_1) = 0 = {\rm Re}(z_2)$.  To compute
$\int_{{\Bbb R}^+ \times {\rm S}^1} \omega[f]$ we may use a single
cell
        \[
        {\rm D} : \quad -\infty < x < +\infty,
        \quad -\pi < y < + \pi ,
        \]
where $x = {\rm Re}(z_1)$ and $y = {\rm Im}(z_2)$.  The boundary
$\partial{\rm D}$ consists of the two lines $y = -\pi$ and $y = \pi$
$(x \in {\Bbb R})$.  Using (\ref{gl11}), paying attention to the
orientation of the boundary, and simplifying terms, one finds the
following explicit expression for the integral of $\omega$:
        \begin{eqnarray}
        \int_{{\Bbb R}^+ \times {\rm S}^1} \omega[f] &=& {1 \over 4\pi} 
        \int_{-\infty}^{\infty} dx \int_{-\pi}^{\pi} dy \ {(x-iy)^2 
        \over \cosh(x-iy) - 1} \partial_{\zeta_1} \partial_{\zeta_2}
        \ f \Big( \exp \pmatrix{x &\zeta_1\cr \zeta_2 &iy\cr} \Big)
        \nonumber \\
        &+& {1\over 2} \int_{-\infty}^{\infty} {{\rm d}x \over 
        \cosh x + 1} \ f \Big( \pmatrix{e^x &0\cr 0 &-1\cr} \Big) .
        \nonumber
        \end{eqnarray}
By construction, this Berezin integral is invariant under left and
right translations $f(g) \mapsto f(g_L g g_R)$.  Evaluation gives
$\int_{{\Bbb R}^+ \times {\rm S}^1} \omega[1] = 1 \not= 0$.  The naive
guess would have been $\int \omega[1] = (2\pi i)^{-1} \int {\rm d}a
\wedge {\rm d}d~\partial_{\beta} \partial_{\gamma} = 0$ due to
$\partial_{\beta} \partial_{\gamma} \cdot 1 = 0$.  Such reasoning is
false because $\int_{{\Bbb R}^+} {\rm d}a = \infty$.

\subsection{Symmetric spaces: a reminder}
\label{sec:sym_spaces}

A Riemannian (globally) symmetric space is a Riemannian manifold $M$
such that every $p \in {M}$ is an isolated fixed point of an
involutive isometry.  (In normal coordinates $x^i$ centered around
$p$, this isometry is given by $x^i \mapsto -x^i$.)  This definition
implies (cf.~\cite{helgason}) that the Riemann curvature tensor is
covariantly constant, so that ``the geometry is the same everywhere''.
The curvature can be positive, negative or zero, and the symmetric
space is said to be of compact, noncompact or Euclidean type
correspondingly.

According to Cartan's complete classification scheme, there
exist ten\footnote{We here do not distinguish between the 
orthogonal groups in even and odd dimension.} large classes
of symmetric spaces.  Apart from some minor modifications the
motivation for which is given presently, these are the entries
of Table 1:
\smallskip

\begin{center}
\begin{tabular}{|c||c|c|c|}\hline
class & {\rm noncompact type} & {\rm compact type} \\ \hline
$A$ & ${\rm Gl}(N,{\Bbb C})/{\rm U}(N)$ & ${\rm U}(N)$ \\
$A$I & ${\rm Gl}(N,{\Bbb R})/{\rm O}(N)$ & ${\rm U}(N)/{\rm O}(N)$ \\
$A$II & ${\rm U}^*(2N)/{\rm Sp}(N)$ & ${\rm U}(2N)/{\rm Sp}(N)$ \\
$A$III & ${\rm U}(p,q)/{\rm U}(p)\times{\rm U}(q)$ &
${\rm U}(p+q)/{\rm U}(p)\times{\rm U}(q)$ \\
$BD$I & ${\rm SO}(p,q)/{\rm SO}(p)\times{\rm SO}(q)$ &
${\rm SO}(p+q)/{\rm SO}(p)\times{\rm SO}(q)$ \\
$C$II & ${\rm Sp}(p,q)/{\rm Sp}(p)\times{\rm Sp}(q)$ &
${\rm Sp}(p+q)/{\rm Sp}(p)\times{\rm Sp}(q)$ \\
$BD$ & ${\rm SO}(N,{\Bbb C})/{\rm SO}(N)$ & ${\rm SO}(N)$ \\
$C$ & ${\rm Sp}(N,{\Bbb C})/{\rm Sp}(N)$ & ${\rm Sp}(N)$ \\
$C$I & ${\rm Sp}(N,{\Bbb R})/{\rm U}(N)$ & ${\rm Sp}(N)/{\rm U}(N)$ \\
$D$III & ${\rm SO}^*(2N)/{\rm U}(N)$ & ${\rm SO}(2N)/{\rm U}(N)$ \\ \hline
\end{tabular} \\
\bigskip
Table 1: the large families of symmetric spaces
\end{center}

\noindent
The difference from the standard table\cite{helgason} is that some of
the entries of Table 1, namely the spaces of type $A$, $A$I and
$A$II, are not irreducible.  They can be made so by dividing out a
factor ${\rm U}(1)$ $({\Bbb R}^+)$ in the compact (resp. noncompact)
cases.  Division by such a factor is analogous to removing the center
of mass motion from a mechanical system with translational invariance.
It turns out that, with a view to superanalytic extensions (cf.
Example 2.4), it is preferable not to insist on irreducibility but
to ``retain the center of mass motion''.

The next subsection introduces super-generalizations of Cartan's
symmetric spaces which have appeared in the theory of mesoscopic 
and disordered single-particle systems and have come to play an
important role in that field.

\subsection{Riemannian symmetric superspaces (definition)}
\label{sec:r_s_s}

Let $G_\Lambda$ be a complex Lie supergroup that is realized as a
group of supermatrices $g = \left( \mymatrix{g_{00} &g_{01}\cr g_{10}
&g_{11}\cr}\right)$, with matrix elements that take values in a
(sufficiently large) parameter Grassmann algebra $\Lambda = \Lambda_0
+ \Lambda_1$.  If ${\cal G}_{\Bbb C} = {\cal G}_{\Bbb C}^0 + {\cal
G}_{\Bbb C}^1$ is the Lie superalgebra of $G_\Lambda$, the Lie algebra
of $G_\Lambda$ is obtained by taking the even part of the tensor
product with $\Lambda$: ${\rm Lie}(G_\Lambda) = \Lambda_0 \otimes
{\cal G}_{\Bbb C}^0 + \Lambda_1 \otimes {\cal G}_{\Bbb C}^1 = (\Lambda
\otimes {\cal G}_{\Bbb C})_0$.  Thus, if $\{e_i,\epsilon_j\}$ is a
homogeneous basis of complex matrices in ${\cal G}_{\Bbb C}$, an
element $X\in {\rm Lie}(G_\Lambda)$ is expressed by $X = z^i e_i +
\zeta^j \epsilon_j$ with $z^i \in \Lambda_0$ and $\zeta^j \in
\Lambda_1$.

Let $\theta : G_\Lambda \to G_\Lambda$ be an involutory automorphism
and let $H_\Lambda \subset G_\Lambda$ be the subgroup fixed by
$\theta$.  The decomposition into even and odd eigenspaces of
$\theta_*: {\rm Lie}(G_\Lambda) \to {\rm Lie}(G_\Lambda)$ is written
${\rm Lie}(G_\Lambda) = {\rm Lie}(H_\Lambda) + {\cal M}_\Lambda$.
This decomposition is orthogonal with respect to the ${\rm Ad}
(G_\Lambda)$-invariant quadratic form ${\rm B} : {\rm Lie}(G_\Lambda)
\times {\rm Lie}(G_\Lambda) \to \Lambda_0$, ${\rm B}(X,Y) := 
{\mathop{\rm STr}\nolimits}XY$.

Both $G_\Lambda$ and $H_\Lambda$ are supermanifolds with underlying
spaces that are Lie groups and are denoted by $G_{\Bbb C}$ and
$H_{\Bbb C}$.  Passing to the coset spaces one obtains a graded
commutative algebra ${\cal H} = {\cal H}_0 + {\cal H}_1$ of
(Grassmann-analytically continued) holomorphic sections of the bundle
$G_\Lambda / H_\Lambda \to G_{\Bbb C} / H_{\Bbb C}$.  These sections
are called (super-)functions (on $G_\Lambda / H_\Lambda$) for short.
In local complex coordinates $z^1, ..., z^p ; \zeta^1, ..., \zeta^q$
they are written $f(z^1, ..., z^p; \zeta^1, ..., \zeta^q) = \sum
f_{i_1 ...  i_n} (z^1,...,z^p) \zeta^{i_1} ... \zeta^{i_n}$ where the
coefficients $f_{i_1...i_n}(z^1,...,z^p)$ take values in $\Lambda$ 
after Grassmann-analytic continuation.  For coordinate-independent
calculations the alternative notation $f(gH_\Lambda)$ or $f(g \cdot
o)$ is used.  In the following $G_{\Bbb C} / H_{\Bbb C}$ is assumed
to be connected. 

Every $X \in {\rm Lie}(G_\Lambda)$ is associated with a vector field 
(or even derivation) $\hat X : {\cal H} \to {\cal H}$ by 
        \begin{equation}
        (\hat X f)(g\cdot o) = {d\over ds}\Big|_{s=0} 
        f(e^{-sX} g \cdot o) .
        \label{killing}
        \end{equation}
Here $e^{sX}g$ means the usual product of supermatrices, and the
function $f(e^{sX} g\cdot o)$ is determined from $f(g\cdot o)$ by
Grassmann-analytic continuation.  The Lie algebra of even derivations of
${\cal H}$ is a left ${\cal H}_0$-module denoted by ${\rm Der}_0 {\cal
H}$. \footnote{Since we always have the option of expanding with
respect to the anticommuting parameters that may be contained in $X$,
no information is lost by not considering the full left ${\cal
H}$-module of {\it super}derivations of ${\cal H}$, cf. the last
paragraph of Sec.~\ref{sec:g_a_c}.}

A notion of supergeometry on $G_\Lambda / H_\Lambda$ is introduced via
a left-invariant tensor field $\langle \bullet , \bullet \rangle :
{\rm Der}_0 {\cal H} \times {\rm Der}_0 {\cal H} \to {\cal H}_0$.  The
details are as follows.  $G_\Lambda$ acts on $G_\Lambda / H_\Lambda$
by left translation, $T_h^* : f(g\cdot o) \mapsto f\bigl( (hg)\cdot
o\bigr)$.  The left-translate $dT_h(\hat X)$ of a vector field $\hat
X$ is defined by the equation $T_h^* ( dT_h(\hat X)f ) = \hat X(T_h^*
f)$, and one requires:
        \[
        T_h^* \langle dT_h (\hat X) , dT_h (\hat Y) \rangle = 
        \langle \hat X , \hat Y \rangle . 
        \]
This equation determines $\langle \bullet , \bullet \rangle$ uniquely
within a multiplicative constant.  For vector fields of the special
form (\ref{killing}) one obtains
        \[
        \langle \hat X , \hat Y \rangle (g\cdot o) = c_0
        \times{\rm B}\left( ({\rm Ad}(g)^{-1} X)_{{\cal M}_\Lambda},
        ({\rm Ad}(g)^{-1} Y)_{{\cal M}_\Lambda} \right) ,
        \]
where the subscript ${\cal M}_\Lambda$ means projection on the odd
eigenspace of $\theta_*$.  Note that since $\left( {\rm Ad}(gh)^{-1} X
\right)_{{\cal M}_\Lambda} = {\rm Ad}(h)^{-1} \left( {\rm Ad}(g)^{-1}
X \right)_{{\cal M}_\Lambda}$ for $h \in {H}_\Lambda$, this is
well-defined as a function on $G_\Lambda / H_\Lambda$.  The
normalization is fixed by choosing $c_0 = 1$.

The metric tensor $\langle \bullet , \bullet \rangle$ induces a
geometry on the ordinary manifold $G_{\Bbb C} / H_{\Bbb C}$ by
restriction (i.e. by setting all anticommuting variables equal to
zero).  Of course, since the groups $G_{\Bbb C}$ and $H_{\Bbb C}$ are
complex, this geometry is never Riemann.  However there exist
submanifolds in $G_{\Bbb C} / H_{\Bbb C}$ which are Riemannian
symmetric spaces and can be constructed by selecting from the
tangent space $T_o(G_{\Bbb C} / H_{\Bbb C})$ a Lie-triple subsystem
${\cal M}$ (i.e. $[{\cal M},[{\cal M},{\cal M}]] \subset {\cal M}$)
such that the quadratic form ${\rm B}$ restricted to ${\cal M}$ is of
definite sign.  It is then not hard to show\cite{helgason} that the
image of ${\cal M}$ under the exponential map $X \mapsto e^X
H_\Lambda$ is Riemann in the geometry given by restriction of $\langle
\bullet , \bullet \rangle$.  Its completion is a symmetric space.

{\it Definition 2.5}: A {\it Riemannian symmetric superspace} is a
pair $(G_\Lambda / H_\Lambda ; {M})$ where $M$ is a maximal Riemannian
submanifold of the base $G_{\Bbb C} / H_{\Bbb C}$.

{\it Remark 2.6}: The merit of this definition is that it avoids any
use of an adjoint (or ``complex conjugation'') of the Grassmann
variables. \kasten

By the complex structure of $G_{\Bbb C} / H_{\Bbb C}$, the tangent
space ${\cal M}_{\Bbb C} := {\rm T}_o(G_{\Bbb C}/H_{\Bbb C})$
decomposes as ${\cal M}_{\Bbb C} = {\cal M} + i {\cal M}$ where ${\cal
M}$ is taken to be the subspace of ${\cal M}_{\Bbb C}$ on which the
quadratic form ${\rm B}$ is strictly positive.  Now observe that,
since an element $g \in G_{\Bbb C}$ is of the form $g = {\rm diag}
(g_{00},g_{11})$, the group $G_{\Bbb C}$ is a Cartesian product of two
factors, and the same is true for $H_{\Bbb C}$.  Hence, $G_{\Bbb
C}/H_{\Bbb C}$ factors as $G_{\Bbb C}/H_{\Bbb C} = {M}_{\Bbb C}^{0}
\times {M}_{\Bbb C}^{1}$, and ${\cal M}$ is a sum of two spaces:
${\cal M} = {\cal M}_{0} \oplus {\cal M}_{1}$, which are orthogonal
with respect to the quadratic form ${\rm B}$.  (It may happen, of
course, that one of these spaces is trivial.)  For $Z \in {\cal M}$,
let the corresponding orthogonal decomposition be written $Z = X + Y$.
Then ${\rm B}$ restricted to ${\cal M}$ is evaluated as
        \[
        {\rm B}(Z,Z) = {\mathop{\rm Tr}\nolimits}_0 X^2 - 
        {\mathop{\rm Tr}\nolimits}_1 Y^2 ,
        \]
where the relative minus sign between traces is due to supersymmetry
(${\mathop{\rm STr}\nolimits} = {\mathop{\rm Tr}\nolimits}_0 -
{\mathop{\rm Tr}\nolimits}_1$).  The positivity of ${\rm B}$ on ${\cal
M}$ is seen to imply $X = X^\dagger$ and $Y = - Y^\dagger$ (the dagger
denotes hermitian conjugation, i.e. transposition in conjunction with
complex conjugation).

Given $G_\Lambda / H_\Lambda$, the condition that ${M}$ be Riemann and
maximal in $G_{\Bbb C}/H_{\Bbb C}$, fixes ${M}$ uniquely up to two
possibilities: either $T_o({M}) = {\cal M}$, or $T_o({M}) = i{\cal
M}$.  In either case, ${M}$ is a product of two factors, ${M} = {M}_0
\times {M}_1$, both of which are Riemannian symmetric spaces.  In the
first case, ${M}_0$ is of noncompact type and ${M}_1$ is of compact
type, while in the second case it is the other way around.  We adopt
the convention of denoting the compact space by $M_{\rm F}$ and the
noncompact one by $M_{\rm B}$.

In view of Cartan's list of symmetric spaces (Table 1), we arrive at
the following table of large families of Riemannian symmetric
superspaces:
\bigskip

\begin{center}
\begin{tabular}{|c||c|c|c|}\hline
class &$G_\Lambda / H_\Lambda$ & ${M}_{\rm B}$ & ${M}_{\rm F}$\\ \hline
$A|A$ &${\rm Gl}(m|n)$ & $A$ & $A$\\
$A{\rm I}|A{\rm II}$ &${\rm Gl}(m|2n) / {\rm Osp}(m|2n)$ & $A$I & $A$II\\
$A{\rm II}|A{\rm I}$ &${\rm Gl}(m|2n) / {\rm Osp}(m|2n)$ & $A$II & $A$I\\
$A{\rm III}|A{\rm III}$ &${\rm Gl}(m_1+m_2|n_1+n_2) / {\rm Gl}(m_1|n_1) 
\times {\rm Gl}(m_2|n_2)$ & $A$III & $A$III\\
$BD|C$ &${\rm Osp}(m|2n)$ & $BD$ & $C$\\
$C|BD$ &${\rm Osp}(m|2n)$ & $C$ & $BD$\\
$C{\rm I}|D{\rm III}$ &${\rm Osp}(2m|2n) / {\rm Gl}(m|n)$ & $C$I & $D$III\\
$D{\rm III}|C{\rm I}$ &${\rm Osp}(2m|2n) / {\rm Gl}(m|n)$ & $D$III & $C$I\\
$BD{\rm I}|C{\rm II}$ &${\rm Osp}(m_1+m_2|2n_1+2n_2) / {\rm Osp}(m_1|2n_1) 
\times{\rm Osp}(m_2|2n_2)$ & $BD$I & $C$II\\
$C{\rm II}|BD{\rm I}$ &${\rm Osp}(m_1+m_2|2n_1+2n_2) / {\rm Osp}(m_1|2n_1) 
\times{\rm Osp}(m_2|2n_2)$ & $C$II & $BD$I\\ \hline
\end{tabular}\\
\bigskip
Table 2: Riemannian symmetric superspaces
\end{center}

\noindent
Although the entries $A|A$, $BD|C$ and $C|BD$ look extraneous because
they are groups rather than coset spaces, they fit in the same
framework by putting by $G_\Lambda = G \times G$ and $\theta( g_1,
g_2) = (g_2,g_1)$, so $H_\Lambda = {\rm diag}(G\times G) \simeq G$ and
$G_\Lambda / H_\Lambda \simeq G$.

As far as applications to random-matrix theory and disordered
single-particle systems are concerned, the most important structure
carried by Riemannian symmetric superspaces is their
$G_\Lambda$-invariant Berezin measure.  Such a measure always exists
by Definition 2.1 and the existence of local coordinates.  To describe
it in explicit terms, one introduces a local coordinate system by the
exponential map ${\cal M}_\Lambda \to G_\Lambda / H_\Lambda$, $Z
\mapsto \exp(Z) H_\Lambda$.  By straightforward generalization
(replace the Jacobian by the Berezinian) of a corresponding
calculation (cf.~\cite{helgason}) for ordinary symmetric spaces, one
obtains for the principal term of the invariant Berezin measure the
expression $DZ \circ J(Z)$ where $DZ = {\rm d}z^1 \wedge ... \wedge
{\rm d}z^p \otimes \partial_{\zeta^1} ... \partial_{\zeta^q}$ denotes
the flat Berezin measure on ${\cal M}_\Lambda$, and if $T_Z : {\cal
M}_\Lambda \to {\cal M}_\Lambda$ is the linear operator defined by
        \[
        T_Z=\sum_{n=0}^\infty{{\rm ad}^{2n}(Z)\over(2n+1){\rm !}} \ ,
        \]
the function $J(Z) = {\mathop{\rm SDet}\nolimits} T_Z$.  (Note
$\sum_{n=0}^\infty x^{2n}/(2n+1){\rm !} = x^{-1}\sinh x$.)  A
universally valid expression for the anomaly in these coordinates is
not available at present.

\section{Supersymmetry applied to the Gaussian Random-Matrix
Ensemble of class C}
\label{sec:III}

The goal of the remainder of this paper will be to demonstrate that
Riemannian symmetric superspaces, as defined in Sec.~\ref{sec:r_s_s},
arise in a compelling way when Gaussian ensemble averages of ratios of
spectral determinants for random matrices are considered in the
large-$N$ limit.  The example to be discussed in detail will be the
Gaussian ensemble defined over the symplectic Lie algebra ${\rm
sp}(N)$, which has recently been identified\cite{az_prl} as a model
for the ergodic limit of normal-superconducting mesoscopic systems
with broken time-reversal symmetry.

\subsection{The supersymmetry method: a simple example}
\label{sec:example}

The pedagogical purpose of this first subsection is to illustrate our
strategy at a simple example\cite{brezin}.  If ${\rm u}(N)$ is the Lie
algebra of the unitary group in $N$ dimensions, consider on $i{\rm
u}(N)$ (the hermitian $N \times N$ matrices) the Gaussian probability
measure with width $v / \sqrt{N}$.  Denoting by $H$ the elements of
$i{\rm u}(N)$ and by $dH$ a Euclidean measure, we write the Gaussian
probability measure in the form $d\mu(H) = \exp(-N{\mathop{\rm
Tr}\nolimits}H^2 / 2v^2) dH$, $\int d\mu(H) = 1$.  This measure is
called the Gaussian Unitary Ensemble (GUE) in random-matrix theory.
The object of illustration will be the average ratio of spectral
determinants,

        \[
        Z(\alpha,\beta) = \int_{i{\rm u}(N)} {\mathop{\rm Det}\nolimits} 
        \left( {H - \beta \over H - \alpha } \right) d\mu(H) ,
        \]
where $\alpha$, $\beta$ are complex numbers and $\alpha$ is not in the
spectrum of $H$.  Given the generating function $Z$, the GUE average 
resolvent is obtained by
        \[
        \int_{i{\rm u}(N)} {\mathop{\rm Tr}\nolimits}(H-z)^{-1} d\mu(H)
        = {\partial \over \partial\alpha} Z(\alpha,\beta) \Big|_{
        \alpha = \beta = z} .
        \]      
We will now show how to compute $Z$ using a formalism that readily
generalizes to more complicated situations.

To avoid the introduction of indices and have a basis-independent
formulation, we choose to interpret $H$ as a self-adjoint endomorphism
$H \in {\rm End}(V)$ of $N$-dimensional complex space $V := {\Bbb
C}^N$ with a hermitian quadratic form $(x,y) \mapsto \langle \bar x ,
y \rangle_V$.

The supersymmetry method starts by introducing ``bosonic space''
$W_{\rm B} = W_0 = {\Bbb C}$ and ``fermionic space'' $W_{\rm F} = W_1
= {\Bbb C}$.  Auxiliary space is the ${\Bbb Z}_2$-graded sum $W = W_{\rm B}
\oplus W_{\rm F} = {\Bbb C}^{1|1}$.  The Cartesian basis of $W$ is
denoted by $e_{\rm B} = (1,0)$ and $e_{\rm F} = (0,1)$.  Let ${\rm
Hom}_\lambda(W,V) := \lambda_0 \otimes {\rm Hom}(W_{\rm B},V) +
\lambda_1 \otimes {\rm Hom}(W_{\rm F},V)$ where $\lambda = \lambda_0 +
\lambda_1$ is the Grassmann algebra with ${\rm dim}_{\Bbb C}{\rm
Hom}(W_{\rm F},V) = N$ generators.  (Grassmann-analytic continuation
will not be needed here.)  ${\rm Hom}_{\tilde\lambda}(V,W)$ is defined
similarly, with another Grassmann algebra $\tilde\lambda$.  The key
idea is to utilize the Gaussian Berezin integral over the
complex-analytic superspace ${\rm Hom}_\lambda(W,V) \times {\rm
Hom}_{\tilde\lambda}(V,W)$.  Let $D(\psi,\tilde\psi)$ (with $\psi \in
{\rm Hom}_\lambda(W,V)$ and $\tilde\psi \in {\rm Hom}_{\tilde\lambda}
(V,W)$) denote a translation-invariant holomorphic Berezin measure on
this linear space.  If $\psi_{\rm B}$ ($\tilde\psi_{\rm B}$) is the
restriction of $\psi$ ($\tilde\psi$) to a map $W_{\rm B} \to V$
(resp. $V \to W_{\rm B}$), fix a Berezin integral $f \mapsto \int
D(\psi,\tilde\psi) f(\psi, \tilde\psi )$ by choosing for the domain of
integration the subspace $M_{\rm r}$ selected by the linear condition
$\tilde\psi_{\rm B} = \psi_{\rm B}^\dagger$ (the adjoint $\psi_{\rm
B}^\dagger : {\Bbb C}^N \to {\Bbb C}$ being defined by
$\overline{\psi_{\rm B}^\dagger z} = \langle \bar z , \psi_{\rm B}
\cdot 1 \rangle_V$).  Because ${\rm Hom}_\lambda(W,V) \times {\rm
Hom}_{\tilde\lambda}(V,W)$ has complex dimension $(2N,2N)$, the
integral $\int D(\psi,\tilde\psi) f(\psi, \tilde\psi)$ does not change
its value when $f$ is replaced by the rescaled function $f^s(\psi,
\tilde\psi) = f(s\psi,s\tilde\psi)$ $(s \in {\Bbb R})$.  Now with 
${\rm End}_0(W) = {\rm End}(W_{\rm B}) \oplus {\rm End}(W_{\rm F})$
and ${\rm End}_1(W) = {\rm Hom}(W_{\rm B},W_{\rm F}) \oplus 
{\rm Hom}(W_{\rm F},W_{\rm B})$, let
        \[
        {\rm End}_\Lambda(W) := \Lambda_0 \otimes {\rm End}_0(W) 
        + \Lambda_1 \otimes {\rm End}_1(W) ,
        \]
where $\Lambda = \Lambda_0 + \Lambda_1$ is the Grassmann algebra with
${\rm dim}_{\Bbb C}{\rm End}_1(W) = 2$ generators, and pick $A \in
{\rm End}(V)$, $B \in {\rm End}_\Lambda(W)$.  $B$ corresponds to what 
is called a $2\times 2$ supermatrix in physics.  An elementary but useful 
result is that, if we normalize $D(\psi,\tilde\psi)$ by $\int D(\psi,
\tilde\psi) \exp \left( -s^2 {\mathop{\rm Tr}\nolimits} \psi\tilde\psi
\right) = 1$, the identity
        \begin{equation}
        \int D(\psi,\tilde\psi) \exp \left( i{\mathop{\rm Tr}\nolimits}_V 
        A\psi\tilde\psi - i{\mathop{\rm STr}\nolimits}_W B \tilde\psi\psi 
        \right) = {\mathop{\rm SDet}\nolimits}_{V \otimes W} 
        (A \otimes 1 - 1 \otimes B)^{-c}
        \label{gaussian_integral}
        \end{equation}
holds with $c = 1$ provided that the integral exists.  (The parameter
$c$ is introduced for later convenience.)  When $A$ and $B$ are
represented by diagonal matrices, verification of (\ref{gaussian_integral}) 
is a simple matter of doing one-dimensional Gaussian integrals.  The
general case follows by the invariance of $D(\psi,\tilde\psi)$ under
unitary transformations of $V$ and ``superrotations'' in $W$. 

Now introduce elements $E_{\rm BB}$ and $E_{\rm FF}$ of ${\rm
End}_0(W)$ by $E_{\rm BB} e_{\rm B} = e_{\rm B}$, $E_{\rm FF}
e_{\rm F} = e_{\rm F}$, and $E_{\rm BB} e_{\rm F} = E_{\rm FF} e_{\rm
B} = 0$.  By setting $A := H$ and $B := \alpha E_{\rm BB} + \beta
E_{\rm FF} =: \omega$, and using
        \[
        {\mathop{\rm SDet}\nolimits}_{V \otimes W} (H \otimes 1 - 
        1 \otimes \omega) = {\mathop{\rm Det}\nolimits}(H - \alpha) / 
        {\mathop{\rm Det}\nolimits}(H - \beta) ,
        \]
we get a Gaussian integral representation of $Z$:
        \begin{eqnarray}
        Z(\omega) := Z(\alpha,\beta) &=& 
        \int {\mathop{\rm SDet}\nolimits}_{V \otimes W} 
        (H \otimes 1 - 1 \otimes \omega)^{-c} d\mu(H)
        \nonumber \\
        &=& \int D(\psi, \tilde\psi) \int \exp \left( i{\mathop{\rm Tr}
        \nolimits}_V H \psi\tilde\psi - i{\mathop{\rm STr}\nolimits}_W 
        \omega \tilde\psi\psi \right) d\mu(H) .
        \label{int_rep_1}
        \end{eqnarray}
In the next step, the ${\rm GUE}$ ensemble average is subjected to the 
following manipulations:
        \begin{eqnarray}
        &&\int \exp (i{\mathop{\rm Tr}\nolimits}H\psi\tilde\psi) d\mu(H)
        \nonumber \\
        &=&\int_{i{\rm u}(N)} \exp \left( i {\mathop{\rm Tr}\nolimits}H 
        \psi\tilde\psi
        - N {\mathop{\rm Tr}\nolimits} H^2 / 2v^2 \right) dH
        \nonumber\\
        &=& \exp - {v^2 \over 2N} {\mathop{\rm Tr}\nolimits}_V 
        (\psi\tilde\psi)^2 = \exp - {v^2 \over 2N} 
        {\mathop{\rm STr}\nolimits}_W (\tilde\psi\psi)^2
        \label{int_rep_2} \\
        &=& \int_{{\Bbb R}\times i{\Bbb R}} DQ 
        \exp \left( i {\mathop{\rm STr}\nolimits} Q\tilde\psi\psi
        - N {\mathop{\rm STr}\nolimits} Q^2 / 2v^2 \right)
        \nonumber \\
        &=:& \int D\mu(Q) \exp (i{\mathop{\rm STr}\nolimits}
        Q\tilde\psi\psi) .
        \nonumber
        \end{eqnarray}
The fourth equality sign decouples the quartic term ${\mathop{\rm STr}
\nolimits}_W (\tilde\psi\psi)^2$ by introducing an auxiliary
integration over $Q \in {\rm End}_\Lambda(W)$.  In order for this Gaussian
integral to converge, the integration domain for the BB-part $Q_{\rm
BB} : W_{\rm B} \to W_{\rm B}$ (FF-part $Q_{\rm FF} : W_{\rm F} \to
W_{\rm F}$) is taken to be the real (resp. imaginary) numbers.  By
using the relations 
(\ref{gaussian_integral},\ref{int_rep_1},\ref{int_rep_2}) we obtain
        \begin{eqnarray}
        Z(\omega) &=& 
        \int D(\psi,\tilde\psi) \left( \int \exp ( i {\mathop{\rm Tr}
        \nolimits}_V H\psi\tilde\psi ) d\mu(H) \right) \exp -i 
        {\mathop{\rm STr}\nolimits}_W \omega \tilde\psi\psi 
        \nonumber\\
        &=& \int D\mu(Q) \ \int D(\psi,\tilde\psi) 
        \exp i{\mathop{\rm Tr}\nolimits}_V \psi (Q - \omega) \tilde\psi 
        \nonumber\\
        &=& \int D\mu(Q) \ {\mathop{\rm SDet}\nolimits}_{V\otimes W} 
        \left( {1_N} \otimes (Q-\omega) \right)^{-c} 
        \label{int_rep_3}\\
        &=& \int D\mu(Q) \ {\mathop{\rm SDet}\nolimits}_W (Q-\omega)^{-N} 
        \nonumber\\
        &=& \int_{{\Bbb R}\times i{\Bbb R}} DQ \ \exp - N 
        {\mathop{\rm STr}\nolimits} 
        \left( Q^2 / 2v^2 + \ln(Q-\omega) \right) .
        \nonumber
        \end{eqnarray}
These steps reduce an integral over the $N\times N$ matrix $H$ to an
integral over the $2 \times 2$ supermatrix $Q$.  The large parameter
$N$ now appears in the exponent of the integrand, so that the
$Q$-integral can be evaluated by a saddle-point approximation that
becomes exact in the limit $N \to \infty$.  By solving the
saddle-point equation $-Q / v^2 = (Q-\omega)^{-1}$ and doing an
elementary calculation, one obtains Wigner's semicircle law for the 
GUE density of states\cite{mehta}:
        \begin{equation}
        \int {\mathop{\rm Tr}\nolimits}\delta(E-H) d\mu(H) = 
        {N \over \pi v} \sqrt{1 - (E/2v)^2} ,
        \label{semicircle}
        \end{equation}
which will be of use later.

\subsection{Definition of the Gaussian Ensemble of type C}
\label{sec:def_C}

Having run through a simple and well-known example, we now treat in
detail a less trivial case where the reduction to a $Q$-integral
representation requires more care.

The ``physical space'' of our model is $V = {\Bbb C}^2 \otimes {\Bbb
C}^N$.  As before, let $x \mapsto \bar x$ denote complex conjugation,
and fix a symmetric quadratic form $\langle \bullet , \bullet
\rangle_V : V \times V \to {\Bbb C}$ such that the corresponding
hermitian quadratic form $\langle \bar x , y \rangle_V = \overline{
\langle \bar y , x \rangle}_V$ is strictly positive. The transpose and
the adjoint of a linear transformation $L \in {\rm End}(V)$ are
defined by $\langle x , L^{\rm T} y \rangle_V = \langle L x , y
\rangle_V$ and $\langle \bar x , L^\dagger y \rangle_V = \langle
\overline{Lx} , y \rangle_V$ as usual.

Consider now the space, $P$, of self-adjoint Hamiltonians $H \in {\rm
End}(V)$ subject to the linear condition
        \begin{equation}
        H = - {\cal C} H^{\rm T} {\cal C}^{-1},
        \label{ph_symmetry}
        \end{equation}
where ${\cal C}$ is skew and ${\cal C}^2 = -1$.  Clearly, $iP$ is
isomorphic to ${\rm sp}(N) = C_N$ (the symplectic Lie algebra in $2N$
dimensions).  Introducing an orthonormal real basis of $V$ we can
represent $H$ by a $2N \times 2N$ matrix.  The explicit form of such a
matrix is
        \[
        H = \pmatrix{a &b \cr b^\dagger &-a^{\rm T} \cr}, \qquad 
        {\rm if} \qquad {\cal C} = \pmatrix{0 &1_N \cr -1_N &0\cr},
        \]
where $a$ $(b)$ is a complex hermitian (resp. symmetric) $N \times N$ 
matrix.  The Gaussian ensemble to be studied is defined by the probability
measure $d\mu(H) = \exp(- N {\mathop{\rm Tr}\nolimits}H^2 / 2v^2) 
dH$, $\int d\mu(H) = 1$.  For any two $A, B \in {\rm End}(V)$,
        \begin{equation}
        \int_{i\times{\rm sp}(N)} {\mathop{\rm Tr}\nolimits} (AH) 
        {\mathop{\rm Tr}\nolimits}(BH)~d\mu(H) 
        = {v^2 \over 2N} {\mathop{\rm Tr}\nolimits} \left( AB - 
        A {\cal C} B^{\rm T} {\cal C}^{-1} \right) .
        \label{correlator}
        \end{equation}
The joint probability density for the eigenvalues of $H$ has been
given in\cite{az_prl}.

The physical motivation for considering a Gaussian random-matrix
ensemble of the above type (type $C$) comes from the fact\cite{az_prl}
that it describes the ergodic limit of mesoscopic
normal-superconducting hybrid systems with time-reversal symmetry
broken by the presence of a weak magnetic field.  To deal with such
systems, the Bogoliubov-deGennes (BdG) independent-quasiparticle
formalism is used.  The first factor in the tensor product $V = {\Bbb
C}^2 \otimes {\Bbb C}^N$ accounts for the BdG particle-hole degree of
freedom, which is introduced for the purpose of treating the pairing
field of the superconductor within the formalism of first
quantization.  The second factor represents the orbital degrees of
freedom of the electron.  $H$ is the Hamiltonian that enters into the
BdG-equations, and the relation (\ref{ph_symmetry}) expresses the
particle-hole symmetry of the BdG-formalism.

Our goal is to compute the following ensemble average:
        \begin{equation}
        Z_n(\alpha_{1},...,\alpha_{n};\beta_{1},...,\beta_{n})
        = \int_{i{\rm sp}(N)} \prod_{i=1}^n {\mathop{\rm Det}\nolimits} 
        \left( {H - \beta_{i} \over H - \alpha_{i} } \right) d\mu(H) .
        \label{gen_func}
        \end{equation}
By the particle-hole symmetry of $H$, $Z_n$ is invariant under a
reversal of sign for any pair $(\alpha_i,\beta_j)$, so no information 
is lost by restricting all $\alpha_i$ to one half of the complex plane.  
For definiteness, we require
        \begin{equation}
        {\mathop{\rm Im}\nolimits} \alpha_i < 0 \quad (i = 1, ..., n) .
        \label{lower_half}
        \end{equation}
All information about the statistical correlations between the
eigenvalues of $H$ can be extracted from $Z_n$.  For example, the
probability that, given there is an eigenvalue at $E_1$, there
exist $n-1$ eigenvalues at $E_2, ..., E_n$ (regardless of the
positions of all other eigenvalues) is equal to
        \begin{eqnarray}
        R_n(E_1,...,E_n) &&= \lim_{\varepsilon\to 0} 
        \left( - \varepsilon \over \pi \right)^n
        \prod_{l=1}^n {\partial\over\partial\alpha_{2l}}
        \Big|_{\alpha_{2l} = E_l - i\varepsilon} 
        \prod_{l=1}^n {\partial\over\partial\alpha_{2l-1}}
        \Big|_{\alpha_{2l-1} = - E_l - i\varepsilon} 
        \nonumber \\
        &&\times Z_{2n}(\alpha_1,...,\alpha_{2n};
        E_1-i\varepsilon , -E_1-i\varepsilon,..., 
        E_n-i\varepsilon, -E_n-i\varepsilon) .
        \label{n_levels}
        \end{eqnarray}
The function $R_n(E_1,...,E_n)$ is called the $n$-level correlation
function in random-matrix theory\cite{mehta}.

\subsection{Symmetries of the auxiliary space}
\label{sec:auxiliary}

To transcribe the supersymmetry method of Sec.~\ref{sec:example} to 
the computation of $Z_n$ (which involves $n$ ratios of spectral
determinants), a simple and natural procedure would be to enlarge the
auxiliary space $W$ by taking the tensor product with ${\Bbb C}^n$.
However, on using the formula
        \[
        \int \exp ( i {\mathop{\rm Tr}\nolimits} H \psi\tilde\psi) 
        d\mu(H) = \exp \left( - {1 \over 2} \int_{i{\rm sp}(N)} ( 
        {\mathop{\rm Tr}\nolimits}H \psi\tilde\psi )^2 d\mu(H) \right),
        \]
one faces the complication that the second moment $\int ( {\rm
Tr} H \psi\tilde\psi )^2 d\mu(H)$ then is a sum of two terms, see the
right-hand side of (\ref{correlator}).  Consequently, one needs {\it
two} decoupling supermatrices $Q$ (one for each term).  Although this
presents no difficulty of a principal nature, it does lead to rather
complicated notations.  An elegant remedy is to modify the definition
of $\psi$ and $\tilde\psi$ so that $\psi\tilde\psi$ shares the
symmetry (\ref{ph_symmetry}) of the BdG-Hamiltonian $H$.  The two
terms then combine into a single one:

        \[
        \int ( {\mathop{\rm Tr}\nolimits} H \psi\tilde\psi )^2 
        d\mu(H) = {v^2 \over N} {\mathop{\rm STr}\nolimits}_W 
        (\tilde\psi\psi)^2 ,
        \]
which can again be decoupled by a single supermatrix $Q$.  To
implement the symmetry (\ref{ph_symmetry}), we proceed as follows.  

We enlarge the auxiliary space $W = W_{\rm B} \oplus W_{\rm F}$ in
some way (left unspecified for the moment) and fix a rule of
supertransposition ${\rm Hom}_\lambda(W,V) \to {\rm Hom}_\lambda
(V,W)$, $\psi \mapsto \psi^{\rm T}$, and ${\rm Hom}_\lambda(V,W) \to {\rm
Hom}_\lambda (W,V)$, $\tilde\psi \mapsto \tilde\psi^{\rm T}$.  Such a rule
obeys $\psi^{\rm TT} = \psi\sigma$ and $\tilde\psi^{\rm TT} = \sigma
\tilde\psi$, where $\sigma \in {\rm End}_0(W)$ is the operator for
superparity, i.e. $\sigma (x+y) = x-y$ for $x+y \in W_{\rm B} \oplus
W_{\rm F} = W$.  It induces a rule of supertransposition ${\rm
End}_\Lambda (W) \to {\rm End}_\Lambda(W)$, $Q \mapsto Q^{\rm T}$ (no
separate symbol is introduced).  Combination with complex conjugation
gives a rule of hermitian conjugation $\dagger : {\rm End}_0(W) \to
{\rm End}_0(W)$.  Now impose on $\psi \in {\rm Hom}_\lambda(W,V)$,
$\tilde\psi \in {\rm Hom}_\lambda(V,W)$ the linear conditions
        \begin{equation}
        \psi = {\cal C} \tilde\psi^{\rm T} \gamma^{-1} , \quad  
        \tilde\psi = - \gamma \psi^{\rm T} {\cal C}^{-1},
        \label{linear_cond_C}
        \end{equation}
with some invertible even element $\gamma$ of ${\rm End}_0(W)$. 
The mutual consistency of these equations requires
        \begin{equation}
        \gamma = \gamma^{\rm T} \sigma .
        \label{consistency}
        \end{equation}
To see that, insert the transpose of the second equation in
(\ref{linear_cond_C}) into the first one.  Using $\psi^{\rm TT} =
\psi\sigma$ you obtain $\psi = - {\cal C}{{\cal C}^{-1}}^{\rm T}
\psi\sigma \gamma^{\rm T}\gamma^{-1}$.  Since ${\cal C}{{\cal
C}^{-1}}^{\rm T} = - 1$ and $\sigma\gamma^{\rm T} = \gamma^{\rm T}
\sigma$, Eq.~(\ref{consistency}) follows.  The consistency condition
can be implemented by taking $W_{\rm B} = W_{\rm F} = {\Bbb C}^2
\otimes {\Bbb C}^n$, see below.  By multiplying the equations
(\ref{linear_cond_C}) we obtain
        \begin{equation}
        \psi\tilde\psi =
        - {\cal C} (\psi\tilde\psi)^{\rm T} {\cal C}^{-1}, \quad
        \tilde\psi\psi = - \gamma (\tilde\psi\psi)^{\rm T}
        \gamma^{-1} .
        \label{psi_symmetries}
        \end{equation}
The first equation is the desired symmetry relation allowing us
to combine terms. To appreciate the consequences of the 
second equation, note that by the fourth step in (\ref{int_rep_2}) 
the symmetries of $\tilde\psi\psi$ get transferred onto $Q$, so 
that the latter is subject to
        \begin{equation}
        Q = - \gamma Q^{\rm T} \gamma^{-1} .
        \label{Q_symmetries}
        \end{equation}
This symmetry reflects that of the BdG-Hamiltonian $H$, see
(\ref{ph_symmetry}).  The linear space ${\rm End}_\Lambda(W)$, when
given a Lie bracket by the commutator, can be identified with ${\rm
gl}(2n|2n) = {\rm Lie}({\rm Gl}(2n|2n))$.  As $\gamma$ is
supersymmetric ($\gamma = \gamma^{\rm T}\sigma$), (\ref{Q_symmetries})
fixes an ${\rm osp}(2n|2n)$-subalgebra.

$\gamma$ is not unique.  For definiteness we choose it as follows.
Let $\{ E_{ij} \}_{i,j=1,...,M}$ be a canonical basis of ${\rm
End}({\Bbb C}^M)$ satisfying $E_{ij} E_{kl} = \delta_{jk} E_{il}$
(here $M = 2$ or $M = n$).  For $M = 2$ define the Pauli spin
operators $\sigma_x = E_{12} + E_{21}$, $\sigma_y = -i E_{12} + i
E_{21}$, and $\sigma_z = E_{11} - E_{22}$.  The usual rule of
supertransposition on ${\rm End}_\Lambda(W)$ is given by
($\mu,\nu=1,2$ and $i,j=1,...,n$)
        \begin{eqnarray}
        \left(  E_{\rm BB} \otimes E_{\mu\nu} \otimes E_{ij} 
        \right)^{\rm T} &=& E_{\rm BB} \otimes E_{\nu\mu} \otimes E_{ji} , 
        \quad
        \left(  E_{\rm BF} \otimes E_{\mu\nu} \otimes E_{ij} 
        \right)^{\rm T} = - E_{\rm FB} \otimes E_{\nu\mu} \otimes E_{ji} , 
        \nonumber \\
        \left(  E_{\rm FB} \otimes E_{\mu\nu} \otimes E_{ij} 
        \right)^{\rm T} &=& E_{\rm BF} \otimes E_{\nu\mu} \otimes E_{ji} , 
        \quad
        \left(  E_{\rm FF} \otimes E_{\mu\nu} \otimes E_{ij} 
        \right)^{\rm T} = E_{\rm FF} \otimes E_{\nu\mu} \otimes E_{ji} .
        \nonumber
        \end{eqnarray}
With these conventions, one possible choice for $\gamma$ is
        \begin{equation}
        \gamma = E_{\rm BB} \otimes \gamma_{\rm B} + E_{\rm FF} \otimes 
        \gamma_{\rm F} \quad {\rm where} \quad
        \gamma_{\rm B} = \sigma_x \otimes 1_n , 
        \quad \gamma_{\rm F} = i\sigma_y \otimes 1_n .
        \label{gamma}
        \end{equation}
This is the choice we make. 

\subsection{Gaussian Berezin integral}

To repeat the steps of Sec.~\ref{sec:example} and derive a
$Q$-integral respresentation for the generating function $Z_n$, we
must first generalize the basic identity (\ref{gaussian_integral}),
whose left-hand side is
        \begin{equation}
        \int D(\psi,\tilde\psi) \ \exp \left( i{\mathop{\rm Tr}\nolimits}_V 
        A\psi\tilde\psi - i{\mathop{\rm STr}\nolimits}_W B 
        \tilde\psi\psi \right) .
        \label{LHS}
        \end{equation}
By (\ref{psi_symmetries}) we have 
        \begin{eqnarray}
        {\mathop{\rm Tr}\nolimits} A\psi\tilde\psi &=& 
        {\mathop{\rm Tr}\nolimits} (\psi\tilde\psi)^{\rm T}
        A^{\rm T} = {\textstyle{1\over 2}} {\mathop{\rm Tr}\nolimits} 
        ( A - {\cal C} A^{\rm T} {\cal C}^{-1} ) \psi\tilde\psi ,
        \nonumber \\
        {\mathop{\rm STr}\nolimits} B\tilde\psi\psi &=& 
        {\mathop{\rm STr}\nolimits} (\tilde\psi\psi)^{\rm T}
        B^{\rm T} = {\textstyle{1\over 2}} {\mathop{\rm Tr}\nolimits} 
        ( B - \gamma B^{\rm T} \gamma^{-1} ) \tilde\psi\psi .
        \nonumber
        \end{eqnarray}
In view of this we demand that $A$ and $B$ satisfy:
        \begin{equation}
        A = - {\cal C} A^{\rm T} {\cal C}^{-1}, \quad
        B = - \gamma B^{\rm T} \gamma^{-1} .
        \label{AB_constraints}
        \end{equation}
When carrying out the calculation (\ref{int_rep_1}-\ref{int_rep_3}) we
need to apply the identity (\ref{gaussian_integral}) twice, the first
time with $A = H$, $B = \omega$, and the second time with $A = 0$, $B
= \omega-Q$.  In order for (\ref{AB_constraints}) to be satisfied with
these identifications, we choose to set
        \[
        \omega = E_{\rm BB} \otimes \sigma_z \otimes
        \sum_{i=1}^n \alpha_i E_{ii} + E_{\rm FF} \otimes \sigma_z
        \otimes \sum_{j=1}^n \beta_j E_{jj} .
        \]
The presence of the factor $\sigma_z = {\rm diag}({+1,-1})$ reverses
the sign of the $\alpha_i$ and $\beta_j$ on that subspace where
$\sigma_z$ acts by multiplication with minus one.  As the imaginary
parts of the $\alpha_i$ control the convergence of the integral, this
sign reversal affects the correct choice of integration domain for
$\psi_{\rm B}$ and $\tilde\psi_{\rm B}$.  To ensure convergence of the
integral (\ref{LHS}), we require ${\mathop{\rm Im}\nolimits}
{\mathop{\rm STr}\nolimits} \omega\tilde\psi\psi \le 0$.  This
inequality is achieved by imposing the condition $\tilde \psi_{\rm B}
= (\sigma_z \otimes 1_n) \psi_{\rm B}^\dagger$, which is compatible
with ${\cal C} = i\sigma_y \otimes 1_N$, $\psi_{\rm B} = {\cal C}
\tilde\psi_{\rm B}^{\rm T} \gamma_{\rm B}^{-1}$, and $\gamma_{\rm B} =
\sigma_x \otimes 1_n$.

{\it Lemma 3.1}: Let $D(\psi,\tilde\psi)$ denote a
translation-invariant holomorphic Berezin measure on the subspace of
${\rm Hom}_\lambda(W,V) \times {\rm Hom}_\lambda(V,W)$ defined by
(\ref{linear_cond_C}).  Fix the integration domain by $\tilde
\psi_{\rm B} = (\sigma_z \otimes 1_n) \psi_{\rm B}^\dagger$, and
normalize $D(\psi,\tilde\psi)$ so that $\int D(\psi,\tilde\psi) \exp
(-s^2 {\mathop{\rm Tr}\nolimits} \psi\tilde\psi) = 1$ $(s\in{\Bbb R})$.  
Then if $A \in
{\rm End}(V)$ and $B \in {\rm End}_\Lambda(W)$ are diagonalizable and
satisfy the linear conditions (\ref{AB_constraints}), the identity
(\ref{gaussian_integral}) holds with $c = 1/2$ provided that the
integral exists.

{\it Proof}:  Assume that $A$ and $B$ are represented by diagonal
matrices
        \[
        A = \sigma_z \otimes \sum_{i=1}^N x_i E_{ii} , \qquad
        B = E_{\rm BB} \otimes \sigma_z \otimes \sum_{j=1}^n z_j E_{jj}
         + E_{\rm FF} \otimes \sigma_z \otimes \sum_{j=1}^n y_j E_{jj} ,
        \]
which conforms with (\ref{AB_constraints}).  The right-hand side of
(\ref{gaussian_integral}) then reduces to
        \begin{equation}
        {\mathop{\rm SDet}\nolimits}_{V\otimes W} ( A \otimes 1
        - 1 \otimes B )^{-1/2} = \prod_{i=1}^N \prod_{j=1}^n
        { (x_i - y_j) (x_i + y_j) \over (x_i - z_j)(x_i + z_j) } \ .
        \label{auxiliary}
        \end{equation}
To evaluate the left-hand side write
        \[
        \psi_{\rm B} = \pmatrix{a &b\cr c &d\cr} , \qquad
        \psi_{\rm F} = \pmatrix{\alpha &\beta\cr \gamma &\delta\cr} ,
        \]
where $a,b,c,d$ ($\alpha, \beta, \gamma, \delta$) are complex
$N \times n$ matrices with commuting (resp. anticommuting) matrix
elements.  The constraint $\tilde\psi = - \gamma \psi^{\rm T} 
{\cal C}^{-1}$ results in
        \[
        \tilde\psi_{\rm B} = \pmatrix{-d^{\rm T} &b^{\rm T}\cr 
        -c^{\rm T} &a^{\rm T} \cr} , \qquad
        \tilde\psi_{\rm F} = \pmatrix{-\delta^{\rm T} &\beta^{\rm T}\cr 
        \gamma^{\rm T} &-\alpha^{\rm T}\cr} ,
        \]
and the reality condition $\tilde\psi_{\rm B} = (\sigma_z \otimes 1_n)
\psi_{\rm B}^\dagger$ means $d = - \bar a$ and $c = \bar b$.  The
exponent of the integrand is expressed by
        \begin{eqnarray}
        {\textstyle{1\over 2}}
        {\mathop{\rm Tr}\nolimits} A \psi \tilde \psi - 
        {\textstyle{1\over 2}}
        {\mathop{\rm Tr}\nolimits} B \tilde\psi \psi    
        = \sum_{i=1}^N \sum_{j=1}^n \Big( &&(x_i-z_j) a_{ij}\bar a_{ij}
        - (x_i + z_j) b_{ij} \bar b_{ij}
        \nonumber \\
        + &&(x_i + y_j) \alpha_{ij}\delta_{ij} - (x_i - y_j)\beta_{ij}
        \gamma_{ij} \Big) .
        \nonumber
        \end{eqnarray}
Doing the Gaussian integrals one gets a result that is identical to
(\ref{auxiliary}), which proves the Lemma for diagonal $A$ and $B$.
The general case follows by the invariance properties of $D(\psi,
\tilde\psi)$.

{\it Remark}:  The condition of diagonalizability can of course be 
weakened but we won't need that here. \kasten
 
To apply Lemma 3.1 to our problem, note
        \[
        {\mathop{\rm SDet}\nolimits}_{V\otimes W} \left( H\otimes 1 - 
        1\otimes\omega
        \right)^{1/2} = \prod_{i=1}^n {\mathop{\rm Det}\nolimits}_V 
        \left( { (H-\alpha_i)(H+\alpha_i) \over (H-\beta_i)(H+\beta_i) }
        \right)^{1/2} = \prod_{i=1}^n {\mathop{\rm Det}\nolimits} 
        \left( H-\alpha_i \over H - \beta_i \right) ,
        \]
where in the second step we used the invariance of the ratio of
determinants under $H \mapsto -H$, which is due to the particle-hole
symmetry $H = - {\cal C} H^{\rm T} {\cal C}^{-1}$.  Moreover, note
        \[
        {\mathop{\rm SDet}\nolimits}_{V\otimes W} \left( 1 \otimes (Q-\omega)
        \right)^{-1/2} = {\mathop{\rm SDet}\nolimits}_W (Q-\omega)^{-N}. 
        \]
The previous calculation (\ref{int_rep_1}-\ref{int_rep_3}) thus {\it
formally} goes through with $c = 1/2$, and $i{\rm sp}(N)$ for $i{\rm
u}(N)$, and we arrive at the following representation of the
generating function:
        \begin{equation}
        Z_n(\omega) = \int DQ \exp -N {\mathop{\rm STr}\nolimits} 
        \left( Q^2 / 2v^2 + \ln(Q-\omega) \right) ,
        \label{fin_eq}
        \end{equation}
where the supermatrix $Q = \left( \mymatrix{Q_{\rm BB} &Q_{\rm BF}\cr
Q_{\rm FB} &Q_{\rm FF}\cr} \right)$ is subject to
(\ref{Q_symmetries}).  To make this rigorous, we have to specify
the integration domain for $Q$ and show that the interchange of the
$(\psi,\tilde\psi)$- and $Q$-integrations is permitted.

\subsection{Choice of integration domain}
\label{sec:domain}

If the steps (\ref{int_rep_1}-\ref{int_rep_3}) are to be valid, we
must arrange for all integrals to be convergent, at least.  This is
easily achieved for $Q_{\rm FF}$, the FF-component of $Q$, but
requires substantial labor for $\psi_{\rm B}$, $\tilde\psi_{\rm B}$
and $Q_{\rm BB}$.  Consider $Q_{\rm FF}$ first. Since $-{\mathop{\rm
STr}\nolimits} Q^2 = - {\mathop{\rm Tr}\nolimits} Q_{\rm BB}^2 +
{\mathop{\rm Tr}\nolimits} Q_{\rm FF}^2 + {\rm nilpotents}$, we want
${\mathop{\rm Tr}\nolimits} Q_{\rm FF}Q_{\rm FF} \le 0$, which leads
us to require that $Q_{\rm FF}$ be antihermitian.  
Combination with (\ref{Q_symmetries}) gives
      \[
      Q_{\rm FF} = - \gamma_{\rm F} Q_{\rm FF}^{\rm T}
      \gamma_{\rm F}^{-1} = - Q_{\rm FF}^\dagger
      \]
where $\gamma_{\rm F} = i\sigma_y \otimes 1_n$, see 
(\ref{gamma}).  The solution space of these equations is 
${\rm sp}(n)$, the symplectic Lie algebra in $2n$ dimensions.  Thus we
choose ${\cal U} := {\rm sp}(n)$ for the integration domain of $Q_{\rm
FF}$, and of course the integration measure is taken to be the flat one.

The choice of integration domain for $Q_{\rm BB}$ is a much more
delicate matter and will occupy us for the remainder of this section.
Recall, first of all, that the convergence of
        \[
        \int D(\psi,\tilde\psi) \exp \left( i {
        \mathop{\rm Tr}\nolimits}H\psi\tilde\psi
        - i {\mathop{\rm STr}\nolimits} \omega\tilde\psi\psi \right) 
        \]
requires taking $\tilde\psi_{\rm B} = \beta \psi_{\rm B}^\dagger$
where $\beta := \sigma_z \otimes 1_n$ cancels the minus signs that
multiply the imaginary parts of the parameters $\alpha_i$ in $\omega$.
To ensure the convergence of
        \[
        \int D(\psi,\tilde\psi) \exp i {\mathop{\rm Tr}\nolimits} \psi
        (Q-\omega) \tilde\psi ,
        \]
one is tempted to choose $Q_{\rm BB}$ in such a way that ${\mathop{\rm
Re}\nolimits} {\mathop{\rm Tr}\nolimits} \psi Q \tilde\psi = 0$.
Unfortunately, when this condition is adopted one gets $Q_{\rm BB} =
\beta Q_{\rm BB}^\dagger \beta$, which causes ${\mathop{\rm
Tr}\nolimits}Q_{\rm BB}^2 = {\mathop{\rm Tr}\nolimits}Q_{\rm BB}\beta
Q_{\rm BB}^\dagger \beta$ to be of {\it indefinite sign}, so that the
integral over $Q$ does not exist.

A way out of this difficulty was first described by Sch\"afer and
Wegner\cite{sw} in a related context.  We are now going to formulate
their prescription in a language that anticipates the geometric structure
emerging in the large-$N$ limit.  To simplify the notation, we put
$Q_{\rm BB} = iZ$.  What we need to do is investigate the expression
        \begin{equation}
        \exp(-N {\mathop{\rm Tr}\nolimits} Q_{\rm BB}^2 / 2v^2 
        + i{\mathop{\rm Tr}\nolimits} Q_{\rm BB}
        \tilde\psi_{\rm B}\psi_{\rm B}) = \exp( N 
        {\mathop{\rm Tr}\nolimits} Z^2 / 2v^2 
        - {\mathop{\rm Tr}\nolimits} Z \tilde\psi_{\rm B}\psi_{\rm B}) .
        \label{stacon}
        \end{equation}
The conditions on $Q_{\rm BB}$ translate into
      \[  
      Z = - \gamma_{\rm B} Z^{\rm T} \gamma_{\rm B}^{-1}
      = - \beta  Z^\dagger \beta^{-1} .
      \]
Because $\gamma_{\rm B} = \sigma_x \otimes 1_n$ is symmetric, the
solution space of the first equation is a complex Lie algebra ${\cal
G}_{\Bbb C} \simeq {\rm so}(2n,{\Bbb C})$.  The matrix representation
of an element $Z \in {\cal G}_{\Bbb C}$ is of the form $\left(
\mymatrix{ A &B\cr C &-A^{\rm T}\cr} \right)$ where $B$ and $C$ are
skew.  The second equation $(Z = -\beta Z^\dagger \beta^{-1})$ means
$A = -A^\dagger$ and $C = B^\dagger$, which fixes a real form ${\cal
G} = {\rm so}^*(2n)$ of ${\cal G}_{\Bbb C} = {\rm so}(2n,{\Bbb C})$.
This real form is {\it noncompact} (i.e. ${\cal G} = {\rm Lie} (G)$
with $G$ a noncompact Lie group), which is what causes all the trouble
and is forcing us to work hard.  Its maximal compact subalgebra ${\cal
K}$ is the set of solutions of $X = \beta X \beta^{-1}$ in ${\cal G}$.
From $X = \left( \mymatrix{ A &0\cr 0 &-A^{\rm T}\cr} \right)$ and $A =
-A^\dagger$ we see that ${\cal K} \simeq {\rm u}(n)$.

To display clearly the general nature of the following construction,
we introduce a symmetric quadratic form ${\rm B} : {\cal G}_{\Bbb C}
\times {\cal G} _{\Bbb C} \to {\Bbb C}$ by ${\rm B}(X,Y) =
{\mathop{\rm Tr}\nolimits} XY$.  The Cartan (orthogonal) decomposition
of ${\cal G}$ with respect to this quadratic form is written ${\cal G}
= {\cal K} \oplus {\cal M}$.  An element $Y$ of ${\cal M}$ satisfies
$Y = - \beta Y\beta^{-1}$.  {}From this in conjunction with the
equation fixing ${\cal K}$ $(X = + \beta X \beta^{-1})$ one deduces the
commutation relations
       \begin{equation}
       [{\cal M},{\cal M}] \subset {\cal K}, \quad
       [{\cal K},{\cal M}] \subset {\cal M}, \quad
       [{\cal K},{\cal K}] \subset {\cal K}.
       \label{comrels}
       \end{equation}
Note that the elements of ${\cal M}$ are hermitian while those of
${\cal K}$ are antihermitian.  We will also encounter the complexified
spaces ${\cal K}_{\Bbb C} = {\cal K} + i {\cal K}$ and ${\cal M}_{\Bbb
C} = {\cal M} + i {\cal M}$.  They, too, are orthogonal with respect
to ${\rm B}$ and satisfy the commutation relations (\ref{comrels}).
The element $\beta = \sigma_z \otimes 1_n$ 
satisfies $\beta = - \gamma_{\rm B} \beta^{\rm T}
\gamma_{\rm B}^{-1}$ and can therefore be regarded as an element of
${\cal G}_{\Bbb C}$.  Moreover, $\beta \in i{\cal K} \subset {\cal 
G}_{\Bbb C}$.

Now we embed ${\cal G} = {\cal K} \oplus {\cal M}$ into ${\cal
G}_{\Bbb C}$ by a map $\phi_b$,
       \begin{eqnarray}
       \phi_b \ : \ {\cal K} \times {\cal M} &&\to {\cal G}_{\Bbb C} ,
       \nonumber \\
       (X,Y) &&\mapsto \phi_b(X,Y) = b \times ( X + e^Y \beta e^{-Y} ) ,
       \nonumber
       \end{eqnarray}
where $b \not= 0$ is some constant that will be specified later. 

{\it Lemma 3.2}: $\phi_b({\cal K}\times{\cal M})$ is an analytic
manifold without boundary, and is diffeomorphic to ${\cal G}$.

{\it Proof}:  Analyticity is clear.  To prove the other properties, we
first establish that $\phi_b$ is injective.  For that purpose, we
write $e^Y \beta e^{-Y} = e^{{\rm ad}(Y)}\beta$ where ${\rm ad}(Y)\beta =
[Y,\beta]$ is the adjoint action on ${\cal G}_{\Bbb C}$. Decomposing
the exponential function according to $\exp = \cosh + \sinh$, we write
$\phi_b = \phi_+ + \phi_-$ where
        \begin{eqnarray}
        \phi_+(X,Y) &=& b \times \left( X + 
        \cosh{\mathop{\rm ad}\nolimits}(Y)\beta\right),
        \nonumber \\
        \phi_-(X,Y) &=& b \times \sinh {\mathop{\rm ad}\nolimits}(Y) 
        \beta. \nonumber
        \end{eqnarray}
{}From the commutation relations (\ref{comrels}) and $\beta \in i{\cal
K}$ we see that $\phi_\pm$ takes values $\phi_+(X,Y) \in {\cal
K}_{\Bbb C}$ and $\phi_-(X,Y) \in {\cal M}_{\Bbb C}$.  Since ${\cal
G}_{\Bbb C} = {\cal K}_{\Bbb C} \oplus {\cal M}_{\Bbb C}$ (direct
sum), injectivity is equivalent to the regularity of the maps $X
\mapsto \phi_+(X,Y)$ (with $Y$ viewed as a parameter) and $Y\mapsto
\phi_-(X,Y)$.  The function $\phi_+(X,\cdot) = X + {\rm const}$ is
obviously regular.  By $Y = Y^\dagger$ the element $Y$ is
diagonalizable with real eigenvalues. The regularity of $\phi_-$ then
follows from $\sinh : {\Bbb R}\to{\Bbb R}$ being monotonic and $Y
\mapsto {\mathop{\rm ad}\nolimits}(Y)\beta$ being regular.  This
completes the proof that $\phi_b$ is injective.  The injectivity of
$\phi_b$ means that $\phi_b({\cal K}\times{\cal M})$ is diffeomorphic
to ${\cal G} = {\cal K} \oplus {\cal M}$.  This in turn means that,
since ${\cal G}$ has no boundary, $\phi_b({\cal K}\times{\cal M})$ has
no boundary either. \kasten

We are now going to demonstrate that $\phi_b({\cal K}\times{\cal M})$
for any $b > 0$ may serve as a mathematically satisfactory domain of
integration for the variable $Z$ in (\ref{stacon}). We begin by
investigating the quadratic form ${\mathop{\rm Tr}\nolimits} Z^2 =
{\rm B}(Z,Z)$ on $\phi_b({\cal K}\times{\cal M})$. For this we set $Z
= Z_+ + Z_-$ with $Z_\pm = \phi_\pm(X,Y)$. Using ${\rm B}(Z_+,Z_-) =
0$ (recall ${\cal K}_{\Bbb C} \bot {\cal M}_{\Bbb C}$), ${\rm
B}({\mathop{\rm ad}\nolimits}(Y)A,B) = - {\rm B}(A,{\mathop{\rm
ad}\nolimits}(Y)B)$ and $\cosh^2 - \sinh^2 = 1$, we obtain
       \[
       {\rm B}(Z,Z)/b^2 = {\rm B}(X,X) + 2 {\rm B}(X,\cosh
       {\mathop{\rm ad}\nolimits}(Y) \beta) + {\rm B}(\beta,\beta) .
       \]
The antihermiticity of $X\in{\cal K}$ gives ${\rm B}(X,X) \le 0$.  In
contrast, $\cosh{\mathop{\rm ad}\nolimits}(Y)\beta \in i{\cal K}$ is
hermitian, so ${\rm B}(X,\cosh{\mathop{\rm ad}\nolimits}(Y)\beta) \in
i{\Bbb R}$.  It follows that $\exp(N{\mathop{\rm Tr}\nolimits} Z^2 /
2v^2) = \exp(N{\mathop{\rm Tr}\nolimits}\phi_b(X,Y)^2
/2v^2)$ is decaying with respect to $X$ and oscillatory
w.r.t. $Y$.

We have not yet made any use of $b > 0$ yet. This inequality
comes into play when the coupling term 
        \[ 
        - {\mathop{\rm Tr}\nolimits} Z \tilde\psi_{\rm B}\psi_{\rm B} 
        = - {\rm B}(Z,\tilde\psi_{\rm B}\psi_{\rm B}) = - b {\rm B}(X, 
        \tilde\psi_{\rm B}\psi_{\rm B}) - b {\rm B}( e^Y \beta
        e^{-Y},\tilde\psi_{\rm B}\psi_{\rm B}) 
        \] 
is considered.  {}From (\ref{psi_symmetries}) and $\tilde\psi_{\rm B}
= \beta\psi_{\rm B}^\dagger$ we see that $\tilde\psi_{\rm B}\psi_{\rm B}$
satisfies
        \[
        \tilde\psi_{\rm B}\psi_{\rm B} = - \gamma_{\rm B}
        (\tilde\psi_{\rm B}\psi_{\rm B})^{\rm T} \gamma_{\rm B}^{-1} 
        = + \beta (\tilde\psi_{\rm B}\psi_{\rm B})
        ^\dagger \beta^{-1} ,
        \]
so $\tilde\psi_{\rm B}\psi_{\rm B} \in i{\cal G}$. Since ${\rm B}$ is
real-valued on ${\cal G}\times{\cal G}$, the term ${\rm B}(X,\tilde
\psi_{\rm B}\psi_{\rm B})$ is purely imaginary. The other term,
        \[
        - b {\rm B}(e^Y \beta e^{-Y},\tilde\psi_{\rm B}\psi_{\rm B}) = 
        - b {\mathop{\rm Tr}\nolimits} (\psi_{\rm B}^{\vphantom{\dagger}} 
        e^{2Y} \psi_{\rm B}^\dagger) \le 0
        \]
is never positive if $b > 0$. Hence the real part of the exponential
in (\ref{stacon}) is negative semidefinite for $Q = iZ \in i\phi_b(
{\cal K}\times{\cal M})$ and $b > 0$.  As a result, the integrals over
$Q$ and $\psi, \tilde\psi$ converge if the integration domain for $Q$
is taken to be $i\phi_b({\cal K}\times{\cal M}) \times{\cal U}$
$(b>0)$.  Because $i\phi_b({\cal K}\times{\cal M})\times {\cal U}$ is
an analytic manifold without boundary and Cauchy's theorem applies, we
may perform the shift of integration variables that is implied by the
fourth equality sign in (\ref{int_rep_2}).  Moreover, the presence of
the nonvanishing imaginary parts of of the parameters $\alpha_i$ in
$\omega$ ensures {\it uniform convergence} of the $(\psi, \tilde
\psi)$-integral with respect to $Q$, so that we may interchange the
order of integration (the second equality sign in (\ref{int_rep_3})).
And finally, any breakdown of diagonalizability of $Q - \omega$ occurs
on a set of measure zero, so that the identity
(\ref{gaussian_integral}) (Lemma 3.1) may be used, and all steps
leading to (\ref{fin_eq}) are rigorous.  In summary, we have proved
the following result.

{\bf Theorem 3.3}:  For $V = {\Bbb C}^2 \otimes {\Bbb C}^N$ and 
$W = {\Bbb C}^{1|1} \otimes {\Bbb C}^2 \otimes {\Bbb C}^n$ define the
generating function
        \begin{eqnarray}
        Z_{n,N}(\omega) &=& \int_{i\times{\rm sp}(N)} 
        {\mathop{\rm SDet}\nolimits}_
        {V\otimes W} \left( H\otimes 1 - 1 \otimes\omega \right)^{-1/2}
        \exp\left(-N {\mathop{\rm Tr}\nolimits}H^2 / 2v^2 \right) dH,
        \nonumber \\
        \omega &=& E_{\rm BB} \otimes \sigma_z \otimes
        \sum_{i=1}^n \alpha_i E_{ii} + E_{\rm FF} \otimes \sigma_z 
        \otimes \sum_{j=1}^n \beta_j E_{jj} 
        \quad ({\mathop{\rm Im}\nolimits}\alpha_i < 0) .
        \nonumber
        \end{eqnarray}
Let $DQ$ denote a translation-invariant holomorphic Berezin measure of
the complex-analytic superspace ${\rm osp}(2n|2n)$.  Then for all $N \in 
{\Bbb N}$, $n \in {\Bbb N}$ and $b > 0$, $DQ$ can be normalized so that 
        \begin{equation}
        Z_{n,N}(\omega) = \int_{i\phi_b({\cal K}\times{\cal M}) \times 
        {\cal U}} DQ \exp -N {\mathop{\rm STr}\nolimits} \left( Q^2 / 
        2v^2 + \ln(Q-\omega) \right) ,
        \label{fin_eq_rig}
        \end{equation}
where ${\cal U} = {\rm sp}(n)$, ${\cal K} \simeq {\rm u}(n)$, ${\cal M}$ 
is determined by ${\cal K} \oplus {\cal M} = {\rm so}^*(2n)$, and
$\phi_b(X,Y) = b \left( X + {\rm Ad}(e^Y)(\sigma_z \otimes 1_n)
\right)$. \kasten

We conclude this subsection with a comment.  In the literature a
parameterization of the form $Q = T P T^{-1}$ (cf.~\cite{ps}) has been
very popular.  In our language, this factorization amounts to choosing
for the integration domain of $Q_{\rm BB}$ the image of $\varphi :
{\cal G} = {\cal K} \oplus {\cal M} \to {\cal G}$, $X + Y \mapsto e^Y
X e^{-Y}$.  This is {\it not} a valid choice as $\varphi({\cal G})$
{\it does have a boundary}, namely the light cone $\{Z|{\rm B}(Z,Z) =
0\}$ in ${\cal G}$, so that shifting of integration variables is not
permitted.  (However, it turns out that the error made becomes
negligible in the limit $N \to \infty$, so that the final results
remain valid if that limit is assumed.)

\subsection{Saddle-point supermanifold}
\label{sec:saddle_points}

The result (\ref{fin_eq_rig}) holds for all $N \in {\Bbb N}$.  We are
now going to use the method of steepest descent to show that in the
limit $N\to\infty$, the integral on the right-hand side reduces to an
integral over a Riemannian symmetric superspace of type $D{\rm
III}|C{\rm I}$.

With our choice of normalization, the mean spacing between the
eigenvalues of $H$ scales as $N^{-1}$ for $N \to \infty$, see
(\ref{semicircle}).  We are most interested in the eigenvalues close
to zero as their statistical properties describe those of the
low-lying Bogoliubov independent-quasiparticle energy levels of
mesoscopic normal-superconducting systems\cite{az_prl}.  To probe
their statistical behavior, what we need to do is keep $\hat\omega =
N\omega/\pi v$ (i.e. $\omega$ scaled by the mean level spacing) fixed
as $N$ goes to infinity.  In this limit $\omega \sim {\cal O} (1/N)$
can be treated as a small perturbation and we may expand $N
{\mathop{\rm STr}\nolimits} \ln (Q-\omega) = N {\mathop{\rm
STr}\nolimits} \ln Q - \pi v {\rm STr}Q^{-1}\hat\omega + {\cal
O}(1/N)$ if $Q^{-1}$ exists.

To evaluate the integral (\ref{fin_eq_rig}) by the method of 
steepest descent, we first look for the critical points of the
function $N F(Q) = N {\mathop{\rm STr}\nolimits} ( Q^2 / 2v^2 + 
\ln Q )$.  These are the solutions of
        \[
        F'(Q) = Q/v^2 + Q^{-1} = 0 ,
        \]
or $Q^2 = -v^2$.  The solution spaces, the so-called
``saddle-point supermanifolds'', are nonlinear subspaces of ${\rm
osp}(2n|2n)$, which can be distinguished by the eigenvalues of $Q$.
Of these supermanifolds, which are the ones to select for the 
steepest-descent evaluation of the integral (\ref{fin_eq_rig})?

To tackle this question, we start out by setting all Grassmann
variables to zero.  The BB-part of the saddle-point manifold(s) is
uniquely determined by the forced choice of integration domain
$i\phi_b({\cal K}\times{\cal M})$ and by analyticity.  This is because
the saddle-point manifold must be deformable (using Cauchy's theorem)
into the integration domain without crossing any of the singularities
of ${\rm SDet}(Q-\omega)^{-N}$; and by inspection one finds that this
condition rules out all saddle-point manifolds except for one, which
is $i\phi_v(0\times{\cal M})$, the subspace of the integration
domain $i\phi_b({\cal K}\times{\cal M})|_{b = v}$ obtained by
dropping from ${\cal G} = {\cal K} \oplus {\cal M}$ the ${\cal
K}$ degrees of freedom (these are the directions of steepest descent).
By an argument given in the proof of Lemma 3.2 we know that
$i\phi_v(0\times{\cal M})$ is diffeomorphic to ${\cal M}$.  On
general grounds the latter is diffeomorphic to a coset space $G/K$ by
the exponential map ${\cal M}\to G/K$, $Y \mapsto e^Y K$; where in the
present case $G = \{ g \in {\rm Gl}(2n,{\Bbb C}) | g = \gamma_{\rm B}
{g^{-1}}^{\rm T} \gamma_{\rm B}^{-1} = \beta {g^{-1}}^\dagger
\beta^{-1} \}$, and $K = \{ k \in G | k = \beta k \beta^{-1} \}$ (on
setting $g = \exp Z$, $k = \exp X$ and linearizing, we recover the
conditions $Z = - \gamma_{\rm B} Z^{\rm T}\gamma_{\rm B}^{-1} = -\beta
Z^\dagger \beta^{-1}$ defining ${\cal G}$ and the condition $X = \beta
X \beta^{-1}$ fixing the subalgebra ${\cal K}$).  We already know
${\cal G} = {\rm so}^*(2n)$ and ${\cal K} \simeq {\rm u}(n)$, so $G =
\exp{\cal G} = {\rm SO}^*(2n)$ and $K = \exp{\cal K} = {\rm U}(n)$.
Because $K$ is a maximal compact subgroup, the coset space $G/K$ is a
Riemannian symmetric space of noncompact type.  In Cartan's notation,
$G/K = {\rm SO}^*(2n) / {\rm U}(n)$ is called type $D$III.  For better
distinction from its FF-analog, we will henceforth denote $G/K$ by
$G/K_{\rm B}$.

We turn to the FF-sector.  Since ${\mathop{\mathop{\rm
SDet}\nolimits}\nolimits}(Q- \omega)^{-N}$ does not have poles but
only has {\it zeroes} as a function of $Q_{\rm FF}$, analyticity
provides {\it no} criterion for selecting any specific solution space
of the saddle-point equation $Q_{\rm FF}^2 = - v^2$.  Instead,
the determining agent now is the limit $N\to\infty$.  From
(\ref{fin_eq_rig}) it is seen that integration over the Gaussian
fluctuations around the saddle-point manifold produces one factor of
$N^{-1}$ ($N^{+1}$) for every commuting (resp. anticommuting)
direction of steepest descent.  Therefore, the limit $N\to\infty$ is
dominated by that saddle-point manifold which has the minimal
transverse (super-)dimension $d_{\rm B}^\bot - d_{\rm F}^\bot$.  A
little thought shows that the transverse dimension is minimized by
choosing $Q_{\rm FF}$ to possess $n$ eigenvalues $+iv$ and $n$
eigenvalues $-iv$. Thus, the dominant saddle-point manifold is
unique and contains the special point $q_0 := iv \beta$ ($\beta
= \sigma_z \otimes 1_n$ now acts in the fermionic subspace).

Recall that the integration domain for $Q_{\rm FF}$ is a compact Lie
algebra ${\cal U} = {\rm sp}(n)$.  The corresponding Lie group $U =
{\rm Sp}(n)$ operates on ${\cal U}$ by the adjoint action
${\mathop{\rm Ad}\nolimits}(u) : {\cal U}\to {\cal U}$, $X \mapsto
uXu^{-1}$.  Because the saddle-point equation $Q_{\rm FF} = -v^2
Q_{\rm FF}^{-1}$ is invariant under this action, the FF-part of the
(dominant) saddle-point manifold can be viewed as the orbit of the
action of ${\mathop{\rm Ad}\nolimits}(U)$ on the special point $q_0
\in {\cal U}$.  Let $K_{\rm F}$ be the stability group of $q_0$, i.e.
$K_{\rm F} = \{ k \in U | k q_0 k^{-1} = q_0 \}$.  By ${\mathop{\rm
Ad}\nolimits}( K_{\rm F}) q_0 = q_0$ the orbit ${\mathop{\rm
Ad}\nolimits}(U) q_0$ is diffeomorphic to the coset space $U/K_{\rm
F}$.  Arguing in the same way as for the BB-sector, one shows that
$K_{\rm F} \simeq K_{\rm B} \simeq {\rm U}(n)$.  Hence $U/K_{\rm F} =
{\rm Sp}(n) / {\rm U}(n)$, which in Cartan's notation is a compact
Riemannian symmetric space of type $C{\rm I}$.

We are finally in a position to construct the full saddle-point {\it
super}manifold.  Recall, first of all, that $Q$ is subject to the
condition $Q = -\gamma Q^{\rm T}\gamma^{-1}$, which defines an
orthosymplectic complex Lie algebra ${\cal G}_\Lambda := {\rm
osp}(2n|2n)$ in ${\rm End}_\Lambda(W)$.  The solution spaces in ${\cal
G}_ \Lambda$ of the equation $Q / v^2 + Q^{-1} = 0$ are
complex-analytic supermanifolds that are invariant under the adjoint
action of the complex Lie supergroup $G_\Lambda := {\rm Osp}(2n|2n)$.
They can be regarded as ${\rm Ad}(G_\Lambda)$-orbits of elements $Q_0
\in {\rm Lie}( G_\Lambda)$ that are solutions of $(Q_0)^2 = -v^2$.
From the above analysis of the BB- and FF-sectors, we know that the
saddle-point supermanifold that dominates in the large-$N$ limit is
obtained by setting $Q_0 = iv\Sigma_z$ where $\Sigma_z = 1_{{\rm
B}|{\rm F}} \otimes \beta = (E_{\rm BB} + E_{\rm FF})\otimes \sigma_z
\otimes 1_n$.  If $H_\Lambda$ is the stability group of $Q_0$, the
orbit ${\rm Ad}(G_\Lambda)Q_0$ is diffeomorphic to the coset space
$G_\Lambda / H_\Lambda$.  From $\gamma\Sigma_z + \Sigma_z \gamma = 0$
and the equation $h\Sigma_z h^{-1} = \Sigma_z$ (or, equivalently, $h =
\Sigma_z h \Sigma_z$) for $h \in H_\Lambda$ one infers $H_\Lambda
\simeq {\rm Gl}(n|n)$.  Hence the unique complex-analytic saddle-point
supermanifold that dominates the large-$N$ limit is $G_\Lambda /
H_\Lambda \simeq {\rm Osp}(2n|2n) / {\rm Gl}(n|n)$.

Turning to the integral (\ref{fin_eq_rig}) we note the relations ${\rm
STr} Q_0^2 = - v^2 {\mathop{\rm STr}\nolimits} 1 = 0$ and $\ln
{\mathop{\rm SDet}\nolimits} Q_0 = \ln 1 = 0$.  
These imply that the function $F(Q) =
{\mathop{\rm STr}\nolimits}(Q^2 / 2v^2 + \ln Q)$ vanishes
identically on ${\rm Ad}(G_\Lambda) Q_0$.  Hence the exponent of the
integrand in (\ref{fin_eq_rig}) restricted to $G_\Lambda / H_\Lambda$
is
        \[
        \pi v {\mathop{\rm STr}\nolimits} Q^{-1} 
        \hat\omega\big|_{G_\Lambda / 
        H_\Lambda} + {\cal O}(1/N) = -i\pi {\rm B}\left(\hat\omega, 
        {\rm Ad}(g)\Sigma_z\right) + {\cal O}(1/N) .  
        \]
To complete the steepest-descent evaluation of (\ref{fin_eq_rig})
we need to Taylor-expand the exponent of the integrand up to second
order and do a Gaussian integral.  By the ${\rm Ad}(G_\Lambda)$-invariance
of the function $N F(Q)$ it is sufficient to do this calculation for
one element of the saddle-point supermanifold, say $Q = Q_0$.  Putting
$Q = Q_0 + Z$ $(Z \in {\cal G}_\Lambda)$ we get
        \[
        N F(Q_0 + Z) = {N \over 2v^2} {\mathop{\rm STr}\nolimits} 
        (Z^2 + Z\Sigma_z Z \Sigma_z) + {\cal O}(Z^3) .
        \]
Now we make the orthogonal decomposition 
${\cal G}_\Lambda = {\rm Lie}(H_\Lambda) + {\cal M}_\Lambda$,
$Z = X + Y$, where $Y = - \Sigma_z Y \Sigma_z$ are the degrees of
freedom tangent to the saddle-point supermanifold, and $X = + \Sigma_z
X \Sigma_z$ are the degrees of freedom transverse to it.  The
translation-invariant Berezin measure $DZ$ of ${\cal G}_\Lambda$
factors as $DZ = DY DX$.  We thus obtain the transverse Gaussian 
integral
        \[
        \int DX \exp \left( - N {\mathop{\rm STr}\nolimits} X^2 / 
        v^2 + {\cal O} (N^0) \right) .
        \]
The integration domain for $X$ is $i{\cal K}_{\rm B} \times {\cal
K}_{\rm F} \simeq i{\rm u}(n) \times {\rm u}(n)$.  By ${\rm dim}\ {\rm
Lie}(H_\Lambda) = (p,q)$ and $p = q$, this integral reduces to a
constant independent of $N$ in the limit $N \to \infty$.

What remains is an integral over the saddle-point supermanifold itself.
Since $DY$ is the local expression of the invariant Berezin measure of
$G_\Lambda / H_\Lambda$ at ${\rm Ad}(e^Y)Q_0|_{Y=0} = Q_0$ we
arrive at the following result.

{\bf Theorem 3.4}: If $Dg_H$ is a suitably normalized invariant
holomorphic Berezin measure of the complex-analytic supermanifold
$G_\Lambda / H_\Lambda \simeq {\rm Osp}(2n|2n) / {\rm Gl}(n|n)$, 
        \begin{equation}
        \lim_{N\to\infty} Z_{n,N}(\pi v \hat\omega / N) = 
        \int_{M_{\rm B} \times M_{\rm F}} Dg_H \exp -i\pi
        {\rm B} \big( \hat\omega , {\rm Ad}(g)\Sigma_z \big)
        \label{fin_eq_infty}
        \end{equation}
where $\Sigma_z = 1_{{\rm B}|{\rm F}} \otimes \sigma_z \otimes 1_n$, 
$M_{\rm B} \simeq {\rm SO}^*(2n) / {\rm U}(n)$, and 
$M_{\rm F} \simeq {\rm Sp}(n) / {\rm U}(n)$. 

{\it Remark 3.5}:  This result expresses the generating function for $N
\to \infty$ as an integral over a Riemannian symmetric superspace of
type $D{\rm III}|C{\rm I}$ (see Tables 1 and 2) with $m = n$.

In \cite{mrz_circular} the $n$-level correlation function $R_n$ is
calculated exactly from (\ref{fin_eq_infty}) for all $n$.

\section{Other symmetry classes}
\label{sec:IV}

There exist 10 known universality classes of ergodic disordered
single-particle systems.  These are the three classic Wigner-Dyson
classes (GOE, GUE, GSE), the three ``chiral'' ones describing a Dirac
particle in a random gauge field (chGUE, chGOE, chGSE), and the four
classes that can be realized in mesoscopic normal-superconducting (NS)
hybrid systems.  In Ref.~\cite{az96} it was noted that there exists a
one-to-one correspondence between these universality classes and the
large families of symmetric spaces (with the exception of the
orthogonal group in odd dimensions).  Specifically, the Gaussian
random-matrix ensemble over the tangent space of the symmetric space
describes the corresponding universality class, in the limit $N \to
\infty$.  In the notation of Table 1 the correspondences are $A
\leftrightarrow {\rm GUE}$, $A{\rm I} \leftrightarrow {\rm GOE}$,
$A{\rm II} \leftrightarrow {\rm GSE}$, $A{\rm III} \leftrightarrow
{\rm chGUE}$, $BD{\rm I} \leftrightarrow {\rm chGOE}$, $C{\rm II}
\leftrightarrow {\rm chGSE}$, and the four NS-classes correspond to
$C$, $D$, $C{\rm I}$, and $D{\rm III}$.

We have shown in detail how to use the supersymmetry method for the
Gaussian ensemble over $C_N = {\rm sp}(N)$, the tangent space of the
symplectic Lie group.  There are nine more ensembles to study.  We will
now briefly run through all these cases, giving only a summary of the
essential changes.

\subsection{Class D}

Recall the definitions given at the beginning of Sec.~\ref{sec:def_C} 
and replace the symplectic unit by ${\cal C} = \sigma_x \otimes 1_N$.
What you get is a Gaussian random-matrix ensemble over $D_N = {\rm so}
(2N)$, the orthogonal Lie algebra in $2N$ dimensions.  The explicit
form of the Hamiltonian is
        \[
        H = \pmatrix{a &b\cr b^\dagger &-a^{\rm T}\cr} 
        \]
where $a$ ($b$) is complex hermitian (resp. skew).  The treatment of
this ensemble closely parallels that of type $C$.  A change first
occurs in the consistency condition for $\gamma$, which now reads
$\gamma = - \gamma^{\rm T}\sigma$ (instead of $\gamma = + \gamma^{\rm
T}\sigma$) by ${\cal C} {{\cal C}^{-1}}^{\rm T} = +1$.  The extra
minus sign can be accommodated by simply exchanging the BB- and
FF-sectors ($\gamma_{\rm B} \leftrightarrow \gamma_{\rm F}$).  The
linear constraint $Q = - \gamma Q^{\rm T}\gamma^{-1}$ again defines an
${\rm osp}(2n|2n)$ Lie algebra, the only difference being that the
BB-sector is now ``symplectic'' while the FF-sector has turned
``orthogonal''.  Everything else goes through as before and we arrive
at the statement of Theorem 3.3  with ${\cal U} \simeq {\rm so}(2n)$,
${\cal K} \simeq {\rm u}(n)$, and ${\cal K} \oplus {\cal M} \simeq 
{\rm sp}(n,{\Bbb R})$.

A novel feature arises in the large-$N$ limit, where instead of one
dominant saddle-point supermanifold there now emerge {\it two}.  One
of them is the orbit with respect to the adjoint action of ${\rm
Osp}(2n|2n)$ on $Q_0 = iv 1_{{\rm B}|{\rm F}} \otimes
\sigma_z \otimes 1_n$ as before, and the other one is the orbit of
        \[
        Q_1 = iv E_{\rm BB} \otimes \sigma_z \otimes 1_n
        + iv E_{\rm FF} \otimes \sigma_z \otimes \left( E_{11} 
        - \sum_{i=2}^{n} E_{ii} \right) .
        \]
(The orbits of $Q_0$ and $Q_1$ are disconnected because the Weyl group
of ${\rm so}(2n)$ is ``too small''.)  Consequently, the right-hand
side of Theorem 3.4 is replaced by a sum of two terms, one for
each of the two saddle-point supermanifolds.  The integral is over a
Riemannian symmetric superspace of type $C{\rm I}|D{\rm III}$ $(m =
n)$ in both cases.

\subsection{Class CI}
\label{sec:class_CI}

Let $V = {\Bbb C}^2 \otimes {\Bbb C}^N$ carry a hermitian inner
product (as always), and consider the space, $P$, of self-adjoint
Hamiltonians $H \in {\rm End}(V)$ of the form
        \[
        H = H^{\rm T} = - {\cal C} H^{\rm T} {\cal C}^{-1}
        = \pmatrix{a &b\cr b &-a\cr} 
        \quad {\rm where} \quad
        {\cal C} = i\sigma_y \otimes 1_N =
        \pmatrix{0 &1_N\cr -1_N &0\cr} .
        \]      
The $N \times N$ matrices $a$ and $b$ are real symmetric.  It is easy
to see\cite{az96} that $P$ is isomorphic to the tangent space of the
symmetric space ${\rm Sp}(N) / {\rm U}(N)$ (type $C$I).  A Gaussian
measure $d\mu(H)$ on $P$ is completely specified by its first two
moments, $\int_P {\mathop{\rm Tr}\nolimits}(AH) d\mu(H) = 0$ and
        \[
        \int_P {\mathop{\rm Tr}\nolimits}(AH) 
        {\mathop{\rm Tr}\nolimits}(BH) d\mu(H) 
        = {v^2 \over 4N} {\mathop{\rm Tr}\nolimits} \left( 
        A(B+B^{\rm T}) - A {\cal C} (B+B^{\rm T}) {\cal C}^{-1} \right) .
        \]
To deal with the random-matrix ensemble defined by this measure,
we take $W = {\Bbb C}^{1|1} \otimes {\Bbb C}^2 \otimes {\Bbb C}^2 
\otimes {\Bbb C}^n$.  Recall $\psi \in {\rm Hom}_\lambda(W,V)$ and 
$\tilde\psi \in {\rm Hom}_{\tilde\lambda}(V,W)$.  The symmetries of $H$ 
are copied to $\psi\tilde\psi$ by imposing the linear conditions
        \[
        \psi = {\cal C} \tilde\psi^{\rm T} \gamma^{-1}, \quad
        \tilde\psi = - \gamma \psi^{\rm T} {\cal C}^{-1} ; \qquad
        \psi = \tilde\psi^{\rm T} \tau^{-1}, \quad
        \tilde\psi = \tau \psi^{\rm T} .
        \]
In order for these conditions to be mutually consistent, $\tau, \gamma
\in {\rm End}_0(W)$ must satisfy
        \[
        \gamma = \gamma^{\rm T} \sigma ,
        \quad \tau = \tau^{\rm T} \sigma ,
        \quad \gamma\tau^{-1} = - \tau\gamma^{-1} .
        \]
Without loss, we take $\gamma$ and $\tau$ to be orthogonal.
The consistency conditions can then be written in the form
        \[
        \gamma^2 = \sigma = \tau^2, \quad
        \gamma\tau + \tau\gamma = 0 .
        \]
If ${\rm Gl}(W) \simeq {\rm Gl}(4n|4n)$ is the Lie supergroup of
regular elements in ${\rm End}_\Lambda(W)$, the equation $\gamma^2 =
\sigma$ in conjunction with $g^{\rm TT} = \sigma g \sigma$ means that
the automorphism $\hat\gamma : {\rm Gl}(W) \to {\rm Gl}(W)$ defined by
$\hat\gamma(g) = \gamma {g^{-1}}^{\rm T} \gamma^{-1}$ is involutory.
The same is true for $\hat\tau$ defined by $\hat\tau(g) = \tau
{g^{-1}}^{\rm T} \tau^{-1}$ and, moreover, $\hat\gamma$ and $\hat\tau$
commute by $\gamma\tau + \tau\gamma = 0$.  For definiteness we take
        \begin{eqnarray}
        \gamma &=& E_{\rm BB} \otimes \gamma_{\rm B} + 
        E_{\rm FF} \otimes \gamma_{\rm F}, \qquad
        \gamma_{\rm B} = \sigma_x \otimes \sigma_z \otimes 1_n ,
        \quad \gamma_{\rm F} = i\sigma_y \otimes 1_2 \otimes 1_n ,
        \nonumber \\
        \tau &=& E_{\rm BB} \otimes \tau_{\rm B} + E_{\rm FF}
        \otimes \tau_{\rm F} , \qquad
        \tau_{\rm B} = 1_2 \otimes \sigma_x \otimes 1_n , \quad
        \tau_{\rm F} = \sigma_z \otimes i\sigma_y \otimes 1_n .
        \nonumber
        \end{eqnarray}
(This choice is consistent with $\tilde\psi_{\rm B} = \beta \psi_{\rm
B}^\dagger$, $\beta = \sigma_z \otimes 1_2 \otimes 1_n$.)  Let 
        \[
        {\cal Q} := \{ Q \in {\rm End}_\Lambda(W) | Q = - \gamma
        Q^{\rm T}\gamma^{-1} = + \tau Q^{\rm T} \tau^{-1} \}
        \] 
be the subspace distinguished by the symmetry properties of
$\tilde\psi\psi$.  The group ${\rm Gl}(W)$ acts on ${\cal Q}$ by $Q
\mapsto gQg^{-1}$.  We now ask what is the subgroup $G_\Lambda$ of
${\rm Gl}(W)$ that leaves the symmetries of $Q$ invariant (the
normalizer of ${\cal Q}$ in ${\rm Gl}(W)$).

{\it Lemma 4.1}: $G_\Lambda$ is isomorphic to ${\rm Osp}(2n|2n)
\times {\rm Osp}(2n|2n)$.

{\it Proof}:  The conditions on $g\in G_\Lambda$ can be phrased as 
follows:
        \[
        \gamma = g \gamma g^{\rm T}, \quad
        \tau = g \tau g^{\rm T} .
        \]
Equivalently, $G_\Lambda$ can be described as the simultaneous ``fixed
point set''\footnote{For lack of a better word we borrow the
terminology from manifolds.  Of course what is meant here is the
supermanifold of solutions in ${\rm Gl}(W)$ of the nonlinear equations
$g = \hat\gamma(g) = \hat\tau(g)$.}  of the involutory automorphisms
$\hat\gamma$ and $\hat\tau$.  We first describe the fixed point set of
${\hat\gamma} \circ {\hat\tau}$, which acts by $({\hat\gamma} \circ 
{\hat\tau})(g) = \varepsilon g \varepsilon^{-1}$ where
$\varepsilon = -i\gamma\tau^{-1}$.  From the explicit expression
$\varepsilon = 1_{{\rm B}|{\rm F}} \otimes \sigma_x \otimes \sigma_y
\otimes 1_n$ we see that $\varepsilon$ has $4n$ eigenvalues equal to
$+1$, $4n$ eigenvalues equal to $-1$, and these are equally
distributed over the bosonic and fermionic subspaces.  Hence the
subgroup of {\rm Gl}(W) fixed by $\hat\gamma \circ \hat\tau$ is
isomorphic to $G_+ \times G_-$ where $G_+ \simeq {\rm Gl}(2n|2n)
\simeq G_-$.  Denote the embedding $G_+ \times G_- \to {\rm Gl}(W)$ by
$\varphi(g_+ , g_-) = g$.  The group $G_\Lambda$ is the fixed point
set of $\hat\tau$ (or, equivalently, of $\hat\gamma$) in $\varphi( G_+
\times G_-)$ ($\hat\tau$ commutes with ${\hat\gamma} \circ {\hat\tau}$
and therefore takes $\varphi ( G_+ \times G_- )$ into itself).  Note
$\varepsilon\tau = - \tau\varepsilon$, ${\varepsilon^{-1}}^{\rm T} = -
\varepsilon$, and for $g \in \varphi( G_+ \times G_- )$ do the
following little calculation:
        \[
        \varepsilon \hat\tau(g) = \varepsilon \tau {g^{-1}}^{\rm T} 
        \tau^{-1} = - \tau\varepsilon {g^{-1}}^{\rm T} \tau^{-1} = 
        \tau {(\varepsilon g)^{-1}}^{\rm T} \tau^{-1} = 
        \hat\tau(\varepsilon g) .
        \]
Combining this with $\varepsilon \varphi(g_+ , g_-) = \varphi(g_+ ,
- g_-)$ one infers that $\hat\tau$ acting on $\varphi(G_+ \times G_-)$
is of the form $\hat\tau \varphi(g_+ , g_-) = \varphi( \hat\tau_+(g_+)
, \hat\tau_- (g_-) )$.  By a short calculation (work in an
eigenbasis of $\varepsilon$) one sees that the involutory
automorphisms $\hat\tau_i : {\rm Gl}(2n|2n) \to {\rm Gl}(2n|2n)$ $(i =
\pm)$ are expressed by $\hat\tau_i(g) = \tau_i {g^{-1}}^{\rm T}
\tau_i^{-1}$ with supersymmetric $\tau_i$ ($\tau_i = \tau_i^{\rm
T}\sigma$).  It follows that $\hat\tau_i$ fixes an orthosymplectic
subgroup of $G_i \simeq {\rm Gl}(2n|2n)$, so $G_\Lambda \simeq {\rm
Osp}(2n|2n) \times {\rm Osp}(2n|2n)$ as claimed.

{\it Corollary 4.2}: The space ${\cal Q}$ is isomorphic to the complement 
of ${\rm osp}(2n|2n) \oplus {\rm osp}(2n|2n)$ in ${\rm osp}(4n|4n)$.

{\it Proof}: The solution space in ${\rm End}_\Lambda(W)$ of $Q =
-\gamma Q^{\rm T} \gamma^{-1}$ is an ${\rm osp}(4n|4n)$ algebra.
Implementing the second condition $Q = + \tau Q^{\rm T} \tau^{-1}$
amounts to removing from ${\rm osp}(4n|4n)$ the subalgebra fixed by $X
= - \tau X^{\rm T} \tau^{-1}$.  By linearization of the conditions $g
= \hat\gamma(g) = \hat\tau(g)$, this subalgebra is identified as ${\rm
Lie}(G_\Lambda) \simeq {\rm osp}(2n|2n) \oplus {\rm osp}(2n|2n)$.
\kasten

The Gaussian integral identity (\ref{gaussian_integral}) continues to
hold, albeit with a different value of $c = 1/4$.  The proof is
essentially the same as before. 

Since ${\cal Q}$ is not a Lie algebra, the description of the 
correct choice of integration domain for the auxiliary variable
$Q$ is more complicated than before.  In the FF-sector we take
${\cal U} := \{ Q_{\rm FF} \in {\cal Q}_{\rm FF} | Q_{\rm FF} = 
- Q_{\rm FF}^\dagger \}$.  By Corollary 4.2, ${\rm sp}(2n) \simeq
\left( {\rm sp}(n) \oplus {\rm sp}(n) \right) \oplus {\cal U}$. 
To deal with the BB-sector we introduce the spaces
        \begin{eqnarray}
        {\cal G} &=& \{ X \in {\rm gl}(2n,{\Bbb C}) |
        X = - \gamma_{\rm B} X^{\rm T} \gamma_{\rm B}^{-1} = 
        - \tau_{\rm B} X^{\rm T} \tau_{\rm B}^{-1} = 
        - \beta X^\dagger \beta^{-1} \} ,
        \nonumber \\
        {\cal M} &=& \{ Y \in {\cal G} | Y = - \beta Y \beta^{-1} \} ,
        \qquad {\cal P}^{\pm} = \{ X \in {\cal Q}_{\rm BB} | 
        X = - \beta X^\dagger \beta^{-1} = \pm \beta X \beta^{-1} \} .
        \nonumber
        \end{eqnarray}
where $\beta = \sigma_z \otimes 1_2 \otimes 1_n$.  The Lie algebra
${\cal G}$ is a noncompact real form of the BB-part of ${\rm Lie}
(G_\Lambda)$.  By $\beta \in i{\cal P}^+$ and the commutation
relations $[{\cal M},{\cal P}^+] \subset {\cal P}^-$ and $[{\cal
M},{\cal P}^-] \subset {\cal P}^+$, we have an embedding
        \begin{eqnarray}
        \phi_b : {\cal P}^+ \times {\cal M} &&\to {\cal Q}_{\rm BB}
        = {\cal P}_{\Bbb C}^+ + {\cal P}_{\Bbb C}^- ,
        \nonumber \\
        (X,Y) &&\mapsto b \times \left( X + e^{{\rm ad}(Y)}\beta
        \right) .
        \nonumber
        \end{eqnarray}
Similar considerations as in Sec.~\ref{sec:domain} show that all
integrals are rendered convergent by the choice of integration domain
$\phi_b ({\cal P}^+ \times {\cal M}) \times {\cal U}$ $(b > 0)$ for
$Q$.  With this choice we again arrive at Theorem 3.3.

The large-$N$ limit is dominated by a single saddle-point
supermanifold, which can be taken as the orbit of $Q_0 = iv
\Sigma_z$ (here $\Sigma_z = 1_{{\rm B}|{\rm F}} \otimes \sigma_z
\otimes 1_2 \otimes 1_n$) under the adjoint action of $G_\Lambda$.
This orbit is diffeomorphic to $G_\Lambda / H_\Lambda$ where
$H_\Lambda = \{ h \in G_\Lambda | h \Sigma_z h^{-1} = \Sigma_z \}$.
The stability group $H_\Lambda$ can equivalently be described as the
fixed point set of $\hat\Sigma_z : G_\Lambda \to G_\Lambda$,
$\hat\Sigma_z(g) = \Sigma_z g \Sigma_z$.  By the relations $\Sigma_z =
\Sigma_z^{\rm T} = - \gamma \Sigma_z \gamma^{-1} = \tau \Sigma_z
\tau^{-1}$ $(\Sigma_z \in {\cal Q})$, the element $\Sigma_z$
anticommutes with $\varepsilon = -i\gamma\tau^{-1}$, and $\hat
\Sigma_z$ commutes with $\hat\gamma \circ \hat\tau$.  These relations
are compatible with the existence of an embedding $\phi : {\rm
Osp}(2n|2n) \times {\rm Osp}(2n|2n) \to {\rm Gl}(W)$ such that $(\hat
\Sigma_z \circ \phi)(g_+ , g_-) = \phi( g_- , g_+ )$.  (Such an
embedding is easily constructed.)  Hence $H_\Lambda \simeq {\rm diag}
\big( {\rm Osp}(2n|2n) \times {\rm Osp}(2n|2n) \big) \simeq {\rm
Osp}(2n|2n)$.  In this way we arrive at Theorem 3.4 with
$G_\Lambda / H_\Lambda \simeq {\rm Osp}(2n|2n)$, and the maximal
Riemannian submanifold $M_{\rm B} \times M_{\rm F}$ where $M_{\rm B}
\simeq {\rm SO}(2n,{\Bbb C}) / {\rm SO}(2n)$ and $M_{\rm F} \simeq
{\rm Sp}(n)$ (type $D|C$).\footnote{Since the orthogonal group here
always appears with an {\it even} dimension, we use the simplified
notation $D|C$, instead of $BD|C$ as in Table 2.}

\subsection{Class DIII}

Consider for $V = {\Bbb C}^2 \otimes {\Bbb C}^2 \otimes {\Bbb C}^N$
the linear space
        \[
        P = \{ H \in {\rm End}(V) | 
        H = H^\dagger = - {\cal C} H^{\rm T} {\cal C}^{-1} = 
        + {\cal T} H^{\rm T} {\cal T}^{-1} \} ,
        \]
where ${\cal C} = \sigma_x \otimes 1_2 \otimes 1_N$ and ${\cal T} =
1_2 \otimes i\sigma_y \otimes 1_N$.  It has been shown\cite{az96} that
$P$ is isomorphic to the tangent space of ${\rm SO}(4N) / {\rm U}(2N)$
(a symmetric space of type $D$III).  Introducing an orthonormal real
basis of $V$, we can represent $H$ by a $4N \times 4N$ matrix.  If
${\cal C}$ and ${\cal T}$ are given by
        \[
        {\cal C} = \pmatrix{    0 &0 &1_N &0    \cr
                                0 &0 &0 &1_N    \cr
                                1_N &0 &0 &0    \cr
                                0 &1_N &0 &0    \cr},
        \qquad
        {\cal T} = \pmatrix{    0 &1_N &0 &0    \cr
                               -1_N &0 &0 &0    \cr
                                0 &0 &0 &1_N    \cr
                                0 &0 &-1_N &0   \cr},
        \]
the explicit form of such a matrix is 
        \[
        H = \pmatrix{   a       &b      &c      &d      \cr
                b^\dagger &a^{\rm T} &-d^{\rm T} &-c^\dagger \cr
                c^\dagger &-d^{\rm T} &-a^{\rm T} &b^\dagger \cr
                        d       &-c     &b      &-a     \cr}
        \]
where all entries are complex $N \times N$ matrices and
$a$, $d$ ($b$, $c$) are hermitian (skew).

The Gaussian random-matrix ensemble on $P$ is defined by the 
Gaussian measure $d\mu(H)$ with vanishing first moment, and
second moment
        \[
        \int_P {\mathop{\rm Tr}\nolimits}(AH) 
        {\mathop{\rm Tr}\nolimits}(BH) d\mu(H) 
        = {v^2 \over 4N} {\mathop{\rm Tr}\nolimits} \left( AB
        - A {\cal C} B^{\rm T} {\cal C}^{-1}
        + A {\cal T} B^{\rm T} {\cal T}^{-1}
        - A ({\cal CT}) B ({\cal CT})^{-1} \right) .
        \]
Given the auxiliary space $W := {\Bbb C}^{1|1} \otimes {\Bbb C}^2
\otimes {\Bbb C}^2 \otimes {\Bbb C}^n$, we impose on $\psi \in {\rm
Hom}_\lambda(W,V)$, $\tilde\psi \in {\rm Hom}_{\tilde\lambda}(V,W)$ 
the linear conditions
        \begin{eqnarray}
        \psi &=& \gamma \tilde\psi^{\rm T} {\cal C}^{-1}, \quad
        \tilde\psi = - {\cal C} \psi^{\rm T} \gamma^{-1} ,
        \nonumber \\
        \psi &=& \tau \tilde\psi^{\rm T} {\cal T}^{-1}, \quad
        \tilde\psi = {\cal T} \psi^{\rm T} \tau^{-1} ,
        \nonumber
        \end{eqnarray}
with some invertible orthogonal elements $\gamma , \tau$ of ${\rm
End}_0(W)$.  Consistency requires $\gamma^2 = -\sigma = \tau^2$
and $\gamma\tau + \tau\gamma = 0$.  A possible choice is
        \begin{eqnarray}
        \gamma &=& \left( 
        E_{\rm BB} \otimes i\sigma_y \otimes 1_2 
        + E_{\rm FF} \otimes \sigma_x \otimes \sigma_z 
        \right) \otimes 1_n ,
        \nonumber \\
        \tau &=& \left( 
        E_{\rm BB} \otimes \sigma_z \otimes i\sigma_y 
        + E_{\rm FF} \otimes 1_2 \otimes \sigma_x 
        \right) \otimes 1_n .
        \nonumber
        \end{eqnarray}
Because this differs from class $C$I only by the exchange
of the bosonic and fermionic subspaces, the following development
closely parallels that for $C$I, and we arrive at another variant
of Theorem 3.3.

The large-$N$ limit is dominated by a pair of complex-analytic
saddle-point supermanifolds, each being isomorphic to ${\rm
Osp}(2n|2n)$.  (The reason why there are two is that ${\rm O}(2n,{\Bbb
C})$ has two connected components.)  The first one is the orbit under
${\rm Ad}(G_\Lambda)$ of $Q_0 = iv 1_{{\rm B} | {\rm F}} \otimes
\sigma_z \otimes 1_2 \otimes 1_n$, and the second one is the orbit of
        \[
        Q_1 = iv \left( E_{\rm BB} \otimes \sigma_z \otimes 1_2
        \otimes 1_n + E_{\rm FF} \otimes \big( 1_2 \otimes \sigma_x
        \otimes E_{11} + \sigma_z \otimes 1_2 \otimes \sum_{i=2}^n E_{ii}
        \big) \right) .
        \]
Both saddle-point supermanifolds are Riemannian symmetric superspaces
of type $C|D$ with dimensionality $m = 2n$ (Table 2). 

\subsection{Class AIII}

The tangent space at the origin of ${\rm U}(p,q) / {\rm U}(p) \times
{\rm U}(q)$ consists of the matrices of the form
        \[
        H = \pmatrix{0 &Z\cr Z^\dagger &0\cr} ,
        \]
where $Z$ is complex and has dimension $p \times q$.  Such matrices
are equivalently described by $H^\dagger = H = - {\cal P} H {\cal
P}^{-1}$ where ${\cal P} = {\rm diag}(1_p , - 1_q)$.  For simplicity,
we will consider only the case $p = q$ (the general case has not yet
been analyzed in the present formalism).  The Gaussian ensemble of
random matrices $H$ is taken to have second moment
        \[
        \int {\mathop{\rm Tr}\nolimits}(AH) 
        {\mathop{\rm Tr}\nolimits}(BH) d\mu(H) 
        = {v^2 \over 2N} {\mathop{\rm Tr}\nolimits} \left( AB
        - A {\cal P} B {\cal P}^{-1} \right).
        \]
The physical space is $V = {\Bbb C}^2 \otimes {\Bbb C}^p$, and the
auxiliary space is $W = {\Bbb C}^{1|1} \otimes {\Bbb C}^2 \otimes
{\Bbb C}^n$.  The definition of $\omega$ is unchanged from class $C$.
To implement the symmetry condition $\psi\tilde\psi = - {\cal P}
\psi\tilde\psi{\cal P}^{-1}$ we set
        \[
        \psi = i{\cal P}\psi\pi^{-1}, \quad
        \tilde\psi = i\pi\tilde\psi{\cal P}^{-1}
        \]
where $\pi = 1_{\rm B|F} \otimes i\sigma_y \otimes 1_n$.  This choice
is consistent with the relation $\tilde\psi_{\rm B} = \beta\psi_{\rm
B}^\dagger$ which ensures convergence of the $(\psi, \tilde \psi
)$-integration.  The auxiliary variable $Q$ ranges over the
complex-analytic superspace
        \[
        {\cal Q} = \{ Q \in {\rm End}_\Lambda(W) | Q = - \pi Q \pi^{-1} \} ,
        \]
and the normalizer of ${\cal Q}$ in ${\rm Gl}(W)$ is:
        \[
        G_\Lambda = \{ g \in {\rm Gl}(W) | g = \pi g \pi^{-1} \}
        \simeq {\rm Gl}(n|n) \times {\rm Gl}(n|n) .
        \]
For the integration domain ${\cal U}$ in the FF-sector we again
take the antihermitian matrices in ${\cal Q}_{\rm FF}$.  In the
BB-sector we set
        \begin{eqnarray}
        {\cal M} &=& \{ Y \in {\rm End}_{\Bbb C}(W_{\rm B}) | 
        Y = \pi Y \pi^{-1} = - \beta Y \beta^{-1} = Y^\dagger \} ,
        \nonumber \\
        {\cal P}^\pm &=& \{ X \in {\rm End}_{\Bbb C}(W_{\rm B}) |
        X = - \pi X \pi^{-1} = \pm \beta X \beta^{-1} = \mp X^\dagger \} .
        \nonumber       
        \end{eqnarray}
The treatment of Sec.~\ref{sec:domain} then goes through as before,
leading again to Theorem 3.3. 

There is a single dominant saddle-point supermanifold, which is the
${\rm Ad}( G_\Lambda )$-orbit of $Q_0 = iv 1_{{\rm B}|{\rm F}}
\otimes \sigma_z \otimes 1_n$ and is diffeomorphic to $G_\Lambda /
H_\Lambda \simeq {\rm Gl}(n|n)$.  The integration domain $M_{\rm B}
\times M_{\rm F}$ is given by $M_{\rm B} \simeq {\rm Gl}(n,{\Bbb C}) /
{\rm U}(n)$ and $M_{\rm F} = {\rm U}(n)$.  The invariant Berezin
measure of this Riemannian symmetric superspace of type $A|A$ was 
discussed for $n = 1$ in Example 2.4.

\subsection{Class BDI}

The form of the random-matrix Hamiltonian $H$ for class $BD$I can be
obtained from the preceding case by taking the $p\times q$ matrix $Z$
to be real.  Put in formulas, $H$ is subject to $H^\dagger = H =
H^{\rm T} = - {\cal P} H {\cal P}^{-1}$.  We again make the restriction
to $p = q$.  The basic correlation law of the Gaussian ensemble is
        \[
        \int {\mathop{\rm Tr}\nolimits}(AH) 
        {\mathop{\rm Tr}\nolimits}(BH) d\mu(H) = {v^2 \over 4N} 
        {\mathop{\rm Tr}\nolimits} \left( A(B+B^{\rm T})
        - A {\cal P} (B + B^{\rm T}) {\cal P}^{-1} \right).
        \]
To accommodate the extra symmetry $H = H^{\rm T}$, auxiliary space is
extended to $W = {\Bbb C}^{1|1} \otimes {\Bbb C}^2 \otimes {\Bbb C}^2
\otimes {\Bbb C}^n$.  The symmetry conditions on $\psi$, $\tilde\psi$
are
        \[
        \psi = i{\cal P}\psi\pi^{-1}, \quad
        \tilde\psi = i\pi\tilde\psi{\cal P}^{-1}; \qquad
        \psi = \tilde\psi^{\rm T} \tau^{-1} , \quad
        \tilde\psi = \tau \psi^{\rm T} ,
        \]
where $\pi = 1_{\rm B|F} \otimes i\sigma_y \otimes 1_2 \otimes 1_n$ and
$\tau = (E_{\rm BB} \otimes 1_2 \otimes \sigma_x + E_{\rm FF} \otimes
1_2 \otimes i\sigma_y ) \otimes 1_n$.  The auxiliary integration space
        \[
        {\cal Q} = \{ Q \in {\rm End}_\Lambda(W) | Q = - \pi Q \pi^{-1} 
        = + \tau Q^{\rm T} \tau^{-1} \}
        \]
has symmetry group (or normalizer)
        \[
        G_\Lambda = \{ g \in {\rm Gl}(W) | g = \pi g \pi^{-1} =
        \tau {g^{-1}}^{\rm T} \tau^{-1} \} \simeq {\rm Gl}(2n|2n) .
        \]
For the integration domain ${\cal U}$ in the FF-sector we once again
take the antihermitian matrices in ${\cal Q}_{\rm FF}$.  In the
BB-sector we set
        \begin{eqnarray}
        {\cal M} &=& \{ Y \in {\rm End}_{\Bbb C}(W_{\rm B}) | 
        Y = \pi Y \pi^{-1} = - \tau Y^{\rm T} \tau^{-1}
        = - \beta Y \beta^{-1} = Y^\dagger \} ,
        \nonumber \\
        {\cal P}^\pm &=& \{ X \in {\rm End}_{\Bbb C}(W_{\rm B}) |
        X = - \pi X \pi^{-1} = + \tau X^{\rm T} \tau^{-1}
        = \pm \beta X \beta^{-1} = \mp X^\dagger \} .
        \nonumber
        \end{eqnarray}
The treatment of Sec.~\ref{sec:domain} then goes through wit modifications
as in Sec.~\ref{sec:class_CI}.   

There is a single dominant saddle-point supermanifold, which is the ${\rm
Ad}( G_\Lambda )$-orbit of $Q_0 = iv 1_{{\rm B}|{\rm F}} \otimes\sigma_z 
\otimes 1_2 \otimes 1_n$ and is diffeomorphic to $G_\Lambda / H_\Lambda
\simeq {\rm Gl}(2n|2n) / {\rm Osp}(2n|2n)$.  The integration domain
$M_{\rm B} \times M_{\rm F}$ is given by $M_{\rm B} \simeq {\rm
Gl}(2n,{\Bbb R}) / {\rm O}(2n)$ and $M_{\rm F} = {\rm U}(2n)/ {\rm
Sp}(n)$.  This is a Riemannian symmetric superspace of type $A{\rm
I}|A{\rm II}$ with $m = 2n$ (Table 2).

\subsection{Class CII}

The tangent space at the origin of ${\rm Sp}(N,N) / {\rm Sp}(N) \times 
{\rm Sp}(N)$ (a noncompact symmetric space of type $C$II) can be
described by the equations
        \[
        H^\dagger = H = - {\cal P} H {\cal P}^{-1} = 
        - {\cal T} H^{\rm T} {\cal T}^{-1} ,
        \]
where ${\cal P} = \sigma_z \otimes 1_2 \otimes 1_N$ and ${\cal T} = 1_2 
\otimes i\sigma_y \otimes 1_N$ (the physical space is $V = {\Bbb C}^2
\otimes {\Bbb C}^2 \otimes {\Bbb C}^N$).  The explicit form of the
matrices is
        \begin{eqnarray}
        H &=& \pmatrix{0 &0 &a &b\cr 0 &0 &-\bar b &\bar a\cr
        a^\dagger &-b^{\rm T} &0 &0\cr b^\dagger &a^{\rm T} &0 &0\cr} ,
        \nonumber \\
        {\rm if} \quad
        {\cal P} &=& \pmatrix{1_N &0 &0 &0\cr 0 &1_N &0 &0\cr
        0 &0 &-1_N &0\cr 0 &0 &0 &-1_N\cr} \quad {\rm and} \quad
        {\cal T} = \pmatrix{0 &1_N &0 &0\cr -1_N &0 &0 &0\cr
        0 &0 &0 &1_N\cr 0 &0 &-1_N &0\cr}, 
        \nonumber
        \end{eqnarray}
where $a$ and $b$ are complex and have dimension $N \times N$.  The
correlation law of the Gaussian random-matrix ensemble of type
$C$II is
        \[
        \int {\mathop{\rm Tr}\nolimits}(AH) 
        {\mathop{\rm Tr}\nolimits}(BH) d\mu(H) 
        = {v^2 \over 4N} {\mathop{\rm Tr}\nolimits} 
        ( A - {\cal P}A{\cal P}^{-1} ) 
        ( B - {\cal T}B^{\rm T}{\cal T}^{-1} ).
        \]
As before, $W = {\Bbb C}^{1|1} \otimes {\Bbb C}^2 \otimes {\Bbb C}^2
\otimes {\Bbb C}^n$.  The symmetry conditions on $\psi$, $\tilde\psi$
are
        \[
        \psi = i{\cal P}\psi\pi^{-1}, \quad
        \tilde\psi = i\pi\tilde\psi{\cal P}^{-1}; \qquad
        \psi = {\cal T}\tilde\psi^{\rm T} \tau^{-1} , \quad
        \tilde\psi = \tau \psi^{\rm T}{\cal T}^{-1} ,
        \]
where $\pi = 1_{\rm B|F} \otimes i\sigma_y \otimes 1_2 \otimes 1_n$ and
$\tau = (E_{\rm BB} \otimes 1_2 \otimes i\sigma_y + E_{\rm FF} \otimes
1_2 \otimes \sigma_x ) \otimes 1_n$.  This differs from class $BD$I
only by the exchange of the bosonic and fermionic subspaces.  Once more 
we arrive at another version of Theorem 3.3. 

There is a only one complex-analytic supermanifold of saddle-points
that dominates for $N \to \infty$.  It is isomorphic to that for class
$BD$I.  The integration domain $M_{\rm B} \times M_{\rm F}$ changes to
$M_{\rm B} \simeq {\rm U}^*(2n) / {\rm Sp}(n)$ and $M_{\rm F} \simeq
{\rm U}(2n) / {\rm O}(2n)$ (not ${\rm U}(2n) / {\rm SO}(2n)$).  This
is a Riemannian symmetric superspace of type $A{\rm II}|A{\rm I}$ with
$m = 2n$ (Table 2).  The group ${\rm U}^*(2n)$ is defined as the
noncompact real subgroup of ${\rm Gl}(2n,{\Bbb C})$ fixed by $g =
{\cal C}\bar g {\cal C}^{-1}$ where ${\cal C} = i\sigma_y \otimes
1_n$.

\subsection{Class A}

This class for $n = 1$ was used to illustrate our general strategy
in Sec.~\ref{sec:example}.  Let us now do the case of arbitrary $n$,
        \[
        Z_n(\alpha_{1},...,\alpha_{n};\beta_{1},...,\beta_{n})
        = \int_{i{\rm u}(N)} \prod_{i=1}^n {\mathop{\rm Det}\nolimits} 
        \left( {H - \beta_{i} \over H - \alpha_{i} } \right) d\mu(H) .
        \]
The classes treated so far $(C, D, C{\rm I}, D{\rm III}, A{\rm III},
BD{\rm I}, C{\rm II})$ all share one feature, namely the existence
of a particle-hole type of symmetry ($H = - {\cal P}H{\cal P}^{-1}$
or $H = - {\cal C}H^{\rm T}{\cal C}^{-1}$), which allows to restrict
all $\alpha_i$ to one half of the complex plane.  Such a symmetry
is absent for the Wigner-Dyson symmetry classes $A$, $A$I, and $A$II,
which results in a somewhat different scenario as it now matters 
how many $\alpha_i$ lie above or below the real axis.  For definiteness
let
        \[
        {\mathop{\rm Im}\nolimits} \alpha_i < 0 \quad (i = 1, ..., n_A), 
        \qquad
        {\mathop{\rm Im}\nolimits} \alpha_j > 0 \quad (j = n_A+1, ..., n) ,
        \]
and set $n_R = n - n_A$. 

Auxiliary space is taken to be $W = {\Bbb C}^{1|1} \otimes {\Bbb C}^n$.
The definition of $\omega$ changes to
        \[
        \omega = E_{\rm BB} \otimes \sum_{i=1}^n \alpha_i E_{ii}
        + E_{\rm FF} \otimes \sum_{j=1}^n \beta_j E_{jj} .
        \]
Recall that the imaginary parts of the $\alpha_i$ steer the convergence
of the $(\psi,\tilde\psi)$-integration.  Since $\omega$ couples to
$\psi, \tilde\psi$ by 
$\exp -i {\mathop{\rm STr}\nolimits}_W \omega \tilde\psi \psi$, 
convergence forces us to take $\tilde\psi_{\rm B} = 
\beta\psi_{\rm B}^\dagger$ where
        \[
        \beta = \sum_{i=1}^{n_A} E_{ii} - \sum_{j = n_A + 1}^n E_{jj}.
        \]
There are no further constraints on $\psi, \tilde\psi$,
or $Q$.  Thus the complex-analytic auxiliary integration
space is ${\cal Q} = {\rm End}_\Lambda(W)$, and $G_\Lambda = {\rm Gl}(W)
\simeq {\rm Gl}(n|n)$.

The integration domain for $Q$ in the FF-sector is taken to be the
antihermitian matrices ${\cal U} = {\rm u}(n)$.  In the BB-sector
we introduce
        \[
        {\cal G} = \{ X \in {\rm gl}(n,{\Bbb C}) |
        X = - \beta X^\dagger \beta^{-1} \}, \qquad
        {\cal K} = \{ X \in {\cal G} | X = \beta X \beta^{-1} \} .
        \]
The Lie algebra ${\cal G}$ is a noncompact real form ${\rm
u}(n_A,n_R)$ of ${\rm gl}(n,{\Bbb C})$, and ${\cal K} = {\rm u}(n_A)
\oplus {\rm u}(n_R)$ is a maximal compact subalgebra.  The space
${\cal M}$ is defined by the Cartan decomposition ${\cal G} = {\cal K}
\oplus {\cal M}$.  The integration domain for $Q_{\rm BB}$ is taken to
be $i\phi_b({\cal K}\times{\cal M})$ where $\phi_b(X,Y) = b (X +
e^{{\rm ad}(Y)} \beta )$ $(b > 0)$.  This gives Theorem 3.3.

By simple power counting, the limit $N \to \infty$ is again dominated
by a single complex-analytic saddle-point supermanifold, which is the
${\rm Ad}(G_\Lambda)$-orbit of $Q_0 = iv 1_{{\rm B}|{\rm F}}
\otimes \beta$.  The stability group $H_\Lambda$ of $Q_0$ is
$H_\Lambda = {\rm Gl}(n_A|n_A) \times {\rm Gl}(n_R|n_R)$, so
        \[
        {\rm Ad}(G_\Lambda)Q_0 \simeq G_\Lambda / H_\Lambda = 
        {\rm Gl}(n|n) / {\rm Gl}(n_A|n_A) \times {\rm Gl}(n_R|n_R).
        \]
The intersection of ${\rm Ad}(G_\Lambda)Q_0$ with $i\phi_v({\cal K}
\times{\cal M})\times{\cal U}$ is $M_{\rm B} \times M_{\rm F}$
where $M_{\rm B} \simeq {\rm U}(n_A , n_R) / {\rm U}(n_A) \times
{\rm U}(n_R)$ and $M_{\rm F} \simeq {\rm U}(n_A + n_R) / {\rm U}(n_A) 
\times {\rm U}(n_R)$.  This is a Riemannian symmetric superspace
of type $A{\rm III}|A{\rm III}$ with $m_1 = n_1 = n_A$ and
$m_2 = n_2 = n_R$ (see Table 2).

\subsection{Class AI}

The tangent space of ${\rm U}(N) / {\rm O}(N)$ is the same as ($i$
times) the real symmetric matrices $H^\dagger = H = H^{\rm T}$.  It
differs from the tangent space of ${\rm SU}(N) / {\rm SO}(N)$, a
symmetric space of type $A{\rm I}$, in an inessential way (just remove
the multiples of the unit matrix).  The Gaussian ensemble over the
real symmetric matrices has its second moment given by
        \[
        \int {\mathop{\rm Tr}\nolimits}(AH) 
        {\mathop{\rm Tr}\nolimits}(BH) d\mu(H) 
        = {v^2 \over 2N} {\mathop{\rm Tr}\nolimits} 
        ( AB + AB^{\rm T} ) .
        \]
This ensemble is related to type $A$ in the same way that type $C$I
is related to type $C$.

To implement the symmetry $H = H^{\rm T}$ we set
$W = {\Bbb C}^{1|1} \otimes {\Bbb C}^2 \otimes {\Bbb C}^n$ and require
$\psi = \tilde\psi^{\rm T} \tau^{-1}$, $\tilde\psi = \tau\psi^{\rm T}$
where $\tau = (E_{\rm BB}\otimes\sigma_x + E_{\rm FF}\otimes i\sigma_y)
\otimes 1_n$.  The auxiliary integration space
        \[
        {\cal Q} = \{ Q\in{\rm End}_\Lambda(W) | 
        Q = \tau Q^{\rm T}\tau^{-1} \}
        \]
has the symmetry group
        \[
        G_\Lambda = \{ g \in {\rm Gl}(W) | g = \tau {g^{-1}}^{\rm T}
        \tau^{-1} \} \simeq {\rm Osp}(2n|2n) .
        \]
The intersection ${\cal U}$ of the FF-sector ${\cal Q}_{\rm FF}$ with 
the  antihermitian matrices is given by ${\rm sp}(n) \oplus {\cal U}
= {\rm u}(2n)$.  In the BB-sector we put
        \begin{eqnarray}
        {\cal M} &=& \{ Y \in {\rm End}_{\Bbb C}(W_{\rm B}) | Y = 
        -\tau Y^{\rm T} \tau^{-1} = -\beta Y\beta^{-1} = Y^\dagger \} ,
        \nonumber \\
        {\cal P}^\pm &=& \{ X \in {\rm End}_{\Bbb C}(W_{\rm B}) |
        X = + \tau X^{\rm T} \tau^{-1}
        = \pm \beta X \beta^{-1} = \mp X^\dagger \} , 
        \nonumber
        \end{eqnarray}
which leads to yet another version of Theorem 3.3.

The large-$N$ limit is controlled by a single complex-analytic
saddle-point supermanifold ${\rm Ad}(G_\Lambda)Q_0 \simeq
G_\Lambda / H_\Lambda$ where $H_\Lambda \simeq {\rm Osp}(2n_A |
2n_A) \times {\rm Osp}(2n_R|2n_R)$ is the stability group of
$Q_0 = iv 1_{{\rm B}|{\rm F}} \otimes \left( \sum_{i=1}^{n_A}
E_{ii} - \sum_{j = n_A+1}^n E_{jj} \right)$.  The intersection
of ${\rm Ad}(G_\Lambda)Q_0$ with the integration domain
$\phi_v({\cal P}^+ \times {\cal M}) \times {\cal U}$ is $M_{\rm B}
\times M_{\rm F}$ where $M_{\rm B} \simeq {\rm SO}(2n_A,2n_R) / {\rm SO}
(2n_A)\times {\rm SO}(2n_R)$ and $M_{\rm F} \simeq {\rm Sp}(n_A + n_R)/
{\rm Sp}(n_A) \times {\rm Sp}(n_R)$.  This is a Riemannian symmetric
superspace of type $BD{\rm I}|C{\rm II}$ (Table 2) with
$m_1 = 2n_1 = 2n_A$ and $m_2 = 2n_2 = 2n_R$. 

\subsection{Class AII}

Finally, the tangent space of ${\rm U}(2N) / {\rm Sp}(N)$ (a symmetric
space of type $A{\rm II}$, except for the substitution ${\rm SU}(2N)
\to {\rm U}(2N)$) can be described as (i times) the subspace of ${\rm
End}({\Bbb C}^2 \otimes {\Bbb C}^N)$ fixed by the linear equations
$H^\dagger = H = {\cal T} H^{\rm T}{\cal T}^{-1}$, ${\cal T} =
i\sigma_y \otimes 1_N$.  The explicit matrix form of $H$ is

        \[
        H = \pmatrix{a &b\cr b^\dagger &a^{\rm T}\cr}
        \]
where $b$ is skew and $a$ is hermitian. 

The conditions $\psi = {\cal T} \tilde\psi^{\rm T}\tau^{-1}$ and
$\tilde\psi = \tau \psi^{\rm T} {\cal T}^{-1}$ are mutually consistent
if, say, $\tau = (E_{\rm BB} \otimes i\sigma_y + E_{\rm FF} \otimes
\sigma_x ) \otimes 1_n$.  The rest of the manipulations leading up
to Theorem 3.3 are the same as for class $A$I, except for the exchange
of the bosonic and fermionic subspaces ($\tau_{\rm B} \leftrightarrow
\tau_{\rm F}$).  The large-$N$ limit is controlled by a single
saddle-point supermanifold $(G_\Lambda / H_\Lambda , M_{\rm B} \times
M_{\rm F})$ where
        \begin{eqnarray}
        G_\Lambda / H_\Lambda &=& {\rm Osp}(2n|2n) / {\rm Osp}(2n_A|2n_A)
        \times {\rm Osp}(2n_R|2n_R) ,
        \nonumber \\
        M_{\rm B} &=& {\rm Sp}(n_A,n_R)/{\rm Sp}(n_A)\times{\rm Sp}(n_R),
        \nonumber \\
        M_{\rm F}&=&{\rm SO}(2n_A+2n_R)/{\rm SO}(2n_A)\times{\rm SO}(2n_R),
        \nonumber
        \end{eqnarray}
which is a Riemannian symmetric superspace of type $C{\rm II}|BD{\rm
I}$ (Table 2) with $m_1 = 2n_1 = 2n_A$ and $m_2 = 2n_2 = 2n_R$.

\section{Summary}
\label{sec:conclusion}

When Dyson realized\cite{dyson70} that the random-matrix ensembles he
had introduced were based on the symmetric spaces of type $A$, $A$I
and $A$II, he wrote: ``The proof of [the] Theorem ... is a mere
verification.  It would be highly desirable to find a more
illuminating proof, in which the appearance of the [final result]
might be related directly to the structure of the symmetric
space...''.  The advent of the supersymmetry method of Efetov and
others has improved the situation lamented by Dyson.  The present
work takes the Gaussian random-matrix ensembles defined over Cartan's
large families of symmetric spaces and, going to the limit of large
matrix dimension, expresses their spectral correlation functions as
integrals over the corresponding Riemannian symmetric superspaces.
These correspondences are summarized in Table 3. \bigskip

\begin{center}
\begin{tabular}{|c|c||c|c|}\hline
RMT &comments &RSS &dimensions\\ \hline
$A$ &Wigner-Dyson (GUE) &$A{\rm III}|A{\rm III}$
&$m_1 = n_1 = n_A$, $m_2 = n_2 = n_R$\\
$A{\rm I}$ &Wigner-Dyson (GOE) &$BD{\rm I}|C{\rm II}$
&$m_1 = 2n_1 = 2n_A$, $m_2 = 2n_2 = 2n_R$\\
$A{\rm II}$ &Wigner-Dyson (GSE) &$C{\rm II}|BD{\rm I}$
&$m_1 = 2n_1 = 2n_A$, $m_2 = 2n_2 = 2n_R$\\ \hline
$A{\rm III}$ ($p = q$) &chiral GUE &$A|A$ &$m = n$\\
$BD{\rm I}$ ($p = q$) &chiral GOE &$A{\rm I}|A{\rm II}$ &$m = n$\\
$C{\rm II}$ ($p = q$) &chiral GSE &$A{\rm II}|A{\rm I}$ &$m = n$\\ \hline
$C$ &NS &$D{\rm III}|C{\rm I}$ &$m = n$\\
$C{\rm I}$ &NS &$D|C$ &$m = 2n$\\
$D$ &NS &$C{\rm I}|D{\rm III}$ &$m = n$\\
$D{\rm III}$ &NS &$C|D$ &$m = 2n$\\ \hline
\end{tabular}\\ \bigskip \end{center}
Table 3: The symmetric-space based random-matrix theories of the first
column map onto the Riemannian symmetric superspaces listed in the
third column.  The notation for the dimensions is taken from Table 2.

The Riemannian symmetric superspaces that appear in Table 3 all have
superdimension $(p,q)$ with $p = q$.  We say that they are ``perfectly
graded'' or ``supersymmetric''.  An interesting question for future
mathematical research is whether our procedure can be optimized by
reducing it to a computation involving no more than the root system of
the symmetric space, thereby obviating the space- and time-consuming
need to distinguish cases.  (Although I have treated all ten cases
separately, it is possible, following Efetov\cite{efetov}, to shorten
the derivation by starting from a large ``master ensemble'' of highest
symmetry and then reducing it by the addition of symmetry-breaking
terms.  I chose not to follow this route as it involves handling
large tensor products, which makes the computations less transparent
and the identification of the spaces involved more difficult.)

The great strength of the supersymmetry method, as compared to other
methods of mesoscopic physics, stems from the fact that it easily
extends beyond the universal random-matrix limit to diffusive and
localized systems.  What one obtains for these more general systems
are field theories of the nonlinear $\sigma$ model type, with fields
that take values in a Riemannian symmetric superspace.  The method
also extends beyond spectral correlations and allows the calculation
of wave function statistics and of transport coefficients such as the 
electrical conductance (see the literature cited in the introduction).

Let me end on a provocative note.  Mathematicians and mathematical
physicists working on supermanifold theory have taken much guidance
from developments in such esoteric subjects as supergravity and
superstring theory.  Wouldn't it be just as worth while to investigate
the beautiful structures outlined in the present paper, whose physical
basis is not speculative but firmly established, and which are of
direct relevance to experiments that are currently being performed in
physics laboratories all over the world?

This research was supported in part by the National Science Foundation
under Grant No. PHY94-07194.

\end{document}